%% file: preprint.tex
\documentclass[iop]{emulateapj-rtx4}
\usepackage{graphicx}
\usepackage{natbib}
\usepackage{lscape}
\bibpunct{(}{)}{;}{a}{}{,}

\newcommand{\mw}{MW}

\newcommand\av{\mbox{$A_V$}}
\newcommand{\msun}{\mbox{M$_\odot$}}
\newcommand{\ee}[1]{\mbox{${} \times 10^{#1}$}}
\newcommand{\eten}[1]{\mbox{$10^{#1}$}}
\newcommand\cmv{\mbox{cm$^{-3}$}}

\newcommand{\sfr }{\mbox{$\dot M_{*}$}}
\newcommand{\fdense }{\mbox{$f_{{\rm dense}}$}}
\newcommand{\mdense }{\mbox{$M_{{\rm dense}}$}}
\newcommand{\sigmasfr}{\mbox{$\Sigma({\rm SFR})$}}
\newcommand{\rhosfr}{\mbox{$\rho({\rm SFR})$}}
\newcommand{\msunmyr}{\mbox{M$_\odot$ Myr$^{-1}$}}
\newcommand{\msunyr}{\mbox{M$_\odot$ {\rm yr}$^{-1}$}}

\newcommand{\kms}{\mbox{km s$^{-1}$}}
\newcommand{\mstar}{\mbox{$M_{\star}$}}
\newcommand{\mcloud}{\mbox{$M_{{\rm cloud}}$}}

\newcommand{\mmol}{\mbox{$M_{{\rm mol}}$}}
\newcommand{\sigmagas}{\mbox{$\Sigma_{{\rm gas}}$}}
\newcommand{\rhogas}{\mbox{$\rho_{{\rm gas}}$}}
\newcommand{\rhoth}{\mbox{$\rho_{{\rm th}}$}}
\newcommand{\sigmamol}{\mbox{$\Sigma_{mol}$}}
\newcommand{\sigmahtwo}{\mbox{$\Sigma_{{\rm H_2}}$}}
\newcommand{\sigmahi}{\mbox{$\Sigma_{\rm HI}$}}
\newcommand{\sigmaatomic}{\mbox{$\Sigma_{\rm{atomic}}$}}
\newcommand{\tff}{\mbox{$t_{{\rm ff}}$}} 
\newcommand{\eff}{\mbox{$\epsilon_{{\rm ff}}$}} 
\newcommand{\fhh}{\mbox{$f_{{\rm H_2}}$}} 
\newcommand{\tcross}{\mbox{$t_{{\rm cross}}$}} 
\newcommand{\torb}{\mbox{$t_{{\rm orb}}$}} 
\newcommand{\tdep}{\mbox{$t_{{\rm dep}}$}} 
\newcommand{\msunpc}{\mbox{M$_\odot$ pc$^{-2}$}}
\newcommand{\sfrkpc}{\mbox{M$_\odot$ yr$^{-1}$ kpc$^{-2}$}}

\newcommand{\mean}[1]{\mbox{$\langle#1\rangle$}} 

\newcommand{\coo}{\mbox{$^{13}$CO}}

\newcommand{\hcop}{\mbox{HCO$^+$}}
\newcommand{\dv}{\mbox{$\Delta$v}}
\newcommand{\jj}[2]{\mbox{$J = #1\rightarrow#2$}}

\slugcomment{}
\shorttitle{Star Formation Relations}
\shortauthors{Evans, Heiderman, Vutisalchavakul}

\begin{document}


\title{Star Formation Relations in Nearby Molecular Clouds}

\author{Neal J. Evans II,
Amanda Heiderman, Nalin Vutisalchavakul}
\affil{The University of Texas at Austin, Department of
Astronomy, 2515 Speedway, Stop C1400,
Austin, TX 78712-1205, USA}
\email{nje@astro.as.utexas.edu}

\input abstract.tex


\input intro.tex

\input models.tex

\input tests.tex

\input disc.tex

\input conclusions.tex

\input ack.tex

\input appendix.tex

\input table1.tex

\bibliographystyle{apj}

\bibliography{robk1,njebib,more}

\input figs.tex

\end{document}

%% file: abstract.tex

\begin{abstract}
We test some ideas for star formation relations against data on
local molecular clouds. On a cloud by cloud basis, the relation
between the surface density of star formation rate and surface density
of gas divided by a free-fall time, calculated from the mean cloud
density, shows no significant correlation.
If a crossing time is substituted for the free-fall time, there is
even less correlation. Within a cloud, the star formation rate
volume and surface densities increase rapidly with the corresponding gas
densities,
faster than predicted by models using the free-fall time defined 
from the local density. A model in which the star formation rate
depends linearly on the mass of gas above a  visual extinction
of 8 mag
describes the data on these clouds, with very 
low dispersion. The data
on regions of very massive star formation, with improved star formation
rates based on free-free emission from ionized gas, also agree with this
 linear relation.
\end{abstract}

%% file: intro.tex

\section{Introduction}\label{intro}

The factors controlling star formation play an important
role in understanding the formation and evolution of galaxies.
Galaxy-scale studies have led to a number of empirical relations
\citep{1998ApJ...498..541K,2008AJ....136.2846B,2004ApJ...606..271G,
2005ApJ...635L.173W}
and theoretical or semi-empirical explanations for these relations
(for a nice review of ideas, see \citealt{2008AJ....136.2782L}).
Most of these observational and theoretical efforts did not consider 
data from studies of star formation in our own Galaxy, for which 
very detailed studies of star formation processes are available.
Recently, there have been attempts to bring these two fields into 
closer communication, as reviewed by Kennicutt and Evans (2012).
In doing so, one must be aware of significant differences and biases,
but the rewards could be substantial.

We use detailed data on the gas and star formation properties
of a sample of nearby ($d < 500$ pc, with one exception at 950 pc) 
clouds to test a number of
suggestions for star formation relations. These clouds have the virtue
of having very complete information, obtained in uniform ways, on their
masses and star formation rates, including resolved studies of these
properties across the face of the clouds. They are the only clouds for
which such detailed studies are available over the whole cloud.
On the other hand, these clouds may not be typical of those in the inner
part of the Galaxy, which may be more characteristic of the regions
generally studied in other galaxies (e.g., \citealt{2013MNRAS.429..987L,
2013arXiv1303.6286K}).

To put the solar neighborhood in context,
we note that the Sun lies in a part of the Galactic disk in which the
gas surface density is dominated by atomic gas ($\sigmaatomic \sim
10$ \msunpc), while the smoothed out surface density of molecular gas
is much less ($\sigmamol \sim 1$ \msunpc) (Dame et al. 2001,
Nakanishi and Sofue 2006). The star formation rate surface density,
$\sigmasfr \sim 3$ \msunpc Gyr$^{-1}$, or $3\ee{-3}$ \sfrkpc\
(Misiriotis 2006).
This \sigmasfr\ is much lower than in the inner parts of the Galaxy
or in many regions studied in other galaxies, but it is close to the
average value over the inner 13.5 kpc of the Galaxy (Kennicutt and Evans 2012).

\subsection{The Sample}\label{sample}

The clouds studied here  are those from the
c2d 
(\citealt{2003PASP..115..965E,2009ApJS..181..321E})
and Gould Belt 
(\citealt{2013AJ....145...94D}, Allen et al. in prep.)
Spitzer legacy programs.
Depending on how they are divided, there are 18-29 clouds in the sample.
In the version we use, there are 29 clouds because we separate clouds
that were mapped separately with their own distances, sizes,
 and velocity dispersions.
These clouds were mapped down to relatively uniform extinction levels
($\av = 2$ mag, except in confused regions) in all the Spitzer bands from
3.6 to 160 \micron. They are also targeted in Herschel surveys covering
60 to 500 \micron\ \citep{2011sca..conf..321A},
and some have been studied in more detail in millimeter continuum emission
(e.g., \citealt{2009ApJ...692..973E})
and molecular lines (e.g., \citealt{2006AJ....131.2921R}).
The properties used in this paper are shown for all 29 clouds in Table 
\ref{tbl-1}, along with median, mean, and standard deviation of the sample.
The size is the effective radius  ($r = \sqrt{A/\pi}$),
 \sfr\ is the star 
formation rate, \mcloud\ is the cloud mass, generally measured down to 
$\av = 2$ mag, \mdense\ is the mass above an extinction contour of 
$\av = 8$ mag, \sigmasfr\ is the surface density of the star formation rate,
\sigmagas\ is the gas surface density averaged over the whole cloud,
\tff\ is the free-fall time calculated from the mean cloud volume density,
\dv\ is the mean linewidth (see Appendix for details), 
and  $\tcross = 2 r/\mean{v}$
is the crossing time for the cloud. 
For the clouds as a whole,
the median, mean,
and standard deviation for the depletion time ($\tdep = \mcloud/\sfr$)
are 106, 201, and 240 Myr, and for the ``efficiency" per free-fall time
($\eff = \tff/\tdep$) are 0.016, 0.018, and 0.013. 
For the ``dense" gas (above the contour of $\av = 8$ mag), 
the median, mean, and standard deviation for \tff\ are 0.53, 0.71, and 0.38 Myr,
for \tdep\ are 45, 47, and 24 Myr, and for \eff\ are 0.019, 0.018, and 0.008.
These values consider the 14 clouds with $\mdense > 0$ and $\sfr >0$.

For this paper, we focus on the gas properties, especially surface and mean
volume densities, derived from extinction maps, and star formation rates,
based on counting young stellar objects (YSOs) identified by their
infrared excess. The basic methods for determining these quantities were
described for the c2d clouds in 
\citet{2009ApJS..181..321E} and \citet{2010ApJ...723.1019H},
and the same techniques were used for the clouds studied in the Gould Belt
project, so we give only a brief summary here.

The Spitzer surveys identified a large fraction of detected sources as
background stars (2\ee4 to 1\ee5 per cloud for the c2d clouds).
Extinctions were determined for each background star by fitting
SEDs to the Spitzer and 2MASS photometry, with stellar photospheric
models from the SSC tool, ``Star-Pet," 
and the extinction law for $R_V = 5.5$ of \citet{2001ApJ...548..296W},
which matches multi-wavelength data well for these
clouds \citep{2009ApJ...690..496C}.
The extinctions to individual stars were averaged over a Gaussian beam
to make extinction and uncertainty maps, based on beam averaging the
uncertainties in the individual extinctions. These maps allow us to probe
extinctions up to $\av \sim 40$ mag. We use maps with angular resolution 
of 270\arcsec\ for all clouds, 
corresponding to 0.34 pc at the mean distances of the clouds (256 pc).
For further details, see \citet{evans07} and updates in 
\citet{2010ApJ...723.1019H}.
The extinction maps were used to calculate the mean extinction within
extinction contours and the uncertainty in that mean propagated from
the uncertainty maps. These averages were used to 
determine mass surface densities in contours of extinction, from which
mean volume densities and total masses could be calculated. We follow
the procedures used by \citet{2010ApJ...723.1019H} in this paper.
For conversion to mass surface density, we use Case A models for
$R_V = 5.5$ dust, given by 
\citet{2001ApJ...548..296W}\footnote{
See \url{http://www.astro.princeton.edu/$\sim$draine/dust/dust.html} for
updates}.
The systematic uncertainties in this conversion are discussed in
the Appendix.

The star formation rates were determined from counting YSOs, multiplying
by the mean mass of stars and dividing by the relevant timescale. 
The identification
of YSOs required careful discrimination against background stars and
galaxies. Star-forming galaxies were the most problematic source
of contaminants.
A combination of color-color and color-magnitude diagrams using both
Spitzer and 2MASS data were used to reject contaminants. For background
galaxies, the SWIRE fields, artifically extincted to match the extinction for
each cloud, were used to characterize the properties of galaxies in
these diagrams. A weighted average over many criteria, optimized to 
remove both stars and galaxies, was adopted, and then examination by
eye was used to remove remaining suspicious objects (about 8\%).
For clouds at low Galactic latitude (e.g., Serpens), background giants
were a further ($\sim 7$\%) source of contamination, and these were 
removed when spectroscopy was available \citep{2009ApJ...691..672O}.
For clouds farther from the plane, contamination was less than 5\%
\citep{2008ApJ...680.1295S}.
The detailed descriptions of this process are in 
\citealt{2007ApJ...663.1149H, evans07, 2009ApJS..181..321E}.
There is an inevitable trade-off between minimizing contamination
and maximizing completeness. The ``official" products for c2d
\citep{2009ApJS..181..321E} and Gould Belt \citep{2013AJ....145...94D}
projects emphasized minimizing contamination. Recently 
\citet{2013ApJS..205....5H} have reanalyzed the c2d clouds using still
more color and magnitude criteria, and they find about $\sim 30$\% more
YSOs. 
If these newer techniques prove reliable, the star formation rates
would increase by about 30\%.

To convert numbers of YSOs to mass of forming stars,
we used $\mean{\mstar} = 0.5$ \msun, based on a fully sampled initial mass
function \citep{2002Sci...295...82K,2003PASP..115..763C,2006SerAJ.172...17N},
consistent with observations of at least one of our clouds;
\citet{2008ApJ...680.1295S} found $\mean{\mstar} = 0.52\pm 0.11$ \msun\
in Cha II.

To get the relevant timescales,
the YSOs were classified into standard SED classes, 
Class I, Flat SED, Class II, and Class III, using spectral indices determined
from all available photometry from 2 to 24 \micron. Further distinction of
Class 0 sources within the larger category of Class I sources used the
bolometric temperature. The timescale relevant to each class was determined
by the number in each class relative to the number of Class II sources, for
which a timescale of 2 Myr was assumed. For tests on scales of the full
cloud, we use all YSOs. For tests on smaller scales, we want to focus on
objects that are still associated with their natal material; for these, we use
only the Class I objects, with a timescale of 0.55 Myr, a slight update from
the value in \citet{2009ApJS..181..321E}, based on the full c2d plus Gould
Belt sample. All references to Class I objects in this paper {\it include}
Class 0 sources as well because the spectral index does not distinguish them.

The low luminosity objects in Classes I and Flat were
most easily confused with extragalactic objects, 
especially for regions of low YSO surface density.
Edge-on disks or Class II objects behind unusally high extinction regions
were less common sources of contamination of the early classes.
\citet{2010ApJ...723.1019H}
removed some of these objects from the sample by requiring detection
of the \jj32 line of \hcop, which traces relatively high densities,
following the findings of 
\citet{2009A&A...498..167V}.
Surveys for \hcop\ toward all Class I and Flat SED objects in the full
c2d and Gould Belt surveys have continued (Heiderman et al. in prep)
but have not been completed. 
Based on the results so far, which
were focused on regions of low extinction,
the number of contaminants is not so large as to affect the Class I
lifetimes significantly, but the fraction of contaminants in
regions of low YSO surface density is about 50\%.
The fraction of Flat SED
sources lacking \hcop\ emission was much higher, 74\% 
\citep{2010ApJ...723.1019H}. 
Since the Flat SED sources
were less clearly associated with their natal material from the start, we
consider only the Class I sources when we focus on smaller scales.
Note that decreasing the number of Class I sources relative to the total
would decrease their timescale, leaving the the star formation rate from
Class I sources essentially unchanged.

The latest version of the YSO sample for the c2d and Gould Belt surveys 
\citep{2013AJ....145...94D}.
has been used here. 
The sample contains 2966 YSO candidates, of which 367 are classified
as Class I sources. \citet{2013AJ....145...94D} further restricted
the sample of ``protostars" by requiring data at submillimeter wavelengths,
resulting in a sample of 230 objects.
While such data was important for getting a luminosity distribution, the
goal of \citet{2013AJ....145...94D}, we do not require it here since
we only wish to count Class I sources.

%% file: models.tex

\section{Models for Star Formation Relations}\label{models}

\subsection{Empirical Relations}\label{empirical}

The most well known star formation relation on the scale of whole
galaxies is the Kennicutt-Schmidt relation 
(\citealt{1959ApJ...129..243S, 1963ApJ...137..758S, 1998ApJ...498..541K}):
\begin{equation}
\sigmasfr = A\, \Sigma{{^N}{_{gas}}}
\label{kslaw}
\end{equation}
where $\sigmagas = \sigmahi + \sigmahtwo$ is the total gas density from
\ion{H}{1} and CO observations, but without correction for helium.
\citet{1998ApJ...498..541K} found a best fit with $N = 1.4\pm 0.15$.
Inclusion of more galaxies and a wider range of galaxy types has not
changed this result \citep{2012ARA&A..50..531K}.
Studies of the radial distribution of star formation within galaxies
have found a similar relation between surface densities, but with more
variation in the best fit for $N$. In addition, a threshold below which
\sigmasfr\ decreases rapidly is found around $\sigmagas = 10$ \msunpc\
(\citealt{2008AJ....136.2846B,2013AJ....146...19L});
this threshold is associated with the transition
between atomic and molecular-dominated ISMs.
When only molecular gas is considered, the values of $N$ tend to be smaller
($N = 1.0-1.4$). Large studies found $N = 1.0\pm 0.2$ for the radial
distributions of star formation and molecular gas surface densities
(\citealt{2008AJ....136.2846B,2013AJ....146...19L}).
The strong, linear correlation with molecular gas has been extended to the
outer regions of galaxies by a line stacking analysis of CO observations
\citep{2011AJ....142...37S}.
That analysis agrees with data in the outer Galaxy that shows star formation
inevitably associated with molecular gas, even in that very 
atomic-dominated part of the Galaxy
\citep{2002ApJ...578..229S}.
These studies are in clear accord that the molecular gas is most
clearly associated with star formation.
However, issues of the conversion of CO observations into molecular column
density, especially for lower metallicity galaxies, introduce significant
uncertainties \citep{2012ARA&A..50..531K,2013ARA&A..51..207B}. 
Since we compare to data in our Galaxy, our conclusions
need not apply for very different metallicities.

Another relation that applies to whole galaxies was derived from
studies of HCN emission (mostly \jj10), which traces denser gas
than does CO 
(\citealt{1988ApJ...334..613S, 2004ApJ...606..271G, 2005ApJ...635L.173W}).
These studies demonstrated a linear relation
between total star formation rate and total mass of dense gas, as
estimated from HCN:
\begin{equation}
\sfr (\msunyr) \sim 1.2\ee{-8} \mdense (\msun),
\label{wueqn}
\end{equation}
where we have taken the specific relation from \citet{2005ApJ...635L.173W}.
This relation had a smaller dispersion than the relation with
total gas or even molecular gas. This relation was consistent with
many studies of star formation in the Galaxy showing that star formation
was highly correlated with relatively dense gas within the molecular cloud
(\citealt{1998ApJ...502..296O, 2007ApJ...666..982E, 2004ApJ...611L..45J,
2010A&A...518L.102A, 1997ApJ...488..277L, 1992ApJ...393L..25L}).

\subsection{Semi-empirical Models}\label{semimodels}

These empirical relations have motivated two apparently conflicting
models. In one, the whole galaxy KS relation is propagated into a
``universal local" relation between volume densities 
\citep{2012ApJ...745...69K}:
\begin{equation}
\rhosfr \propto  \rhogas^x
\end{equation}
The idea is that the amount of gas divided by the free fall time should
be reflected in the star formation rate, and $\tff \propto \rho^{-0.5}$
leading to $x = 1.5$, equal within uncertainties to the exponent in the KS
relation \citep{2005ApJ...630..250K,2007ApJ...669..289K,2008ApJ...684..996N}.
In more definite form, the theoretical relation is given by
\begin{equation}
\rhosfr = f_{{\rm H_2}} \eff \rhogas/\tff,
\end{equation}
where \fhh\ is the fraction of the mass in molecular form  and \eff\
is the ``efficiency" per free-fall time (\tff)
(equation 1 of \citealt{2012ApJ...745...69K}).
Since we focus on molecular gas, we set $ \fhh = 1$.
With 
\begin{equation}
\tff = \sqrt{3 \pi/32 G \rhogas} = 
8.08 {\rm Myr} \rhogas^{-0.5}(\msun {\rm pc^{-3}}),
\end{equation}
we have in cgs units
\begin{equation}
\rhosfr = 4.76\ee{-4} \eff \rhogas^{1.5}
\end{equation}
or, in more convenient units
\begin{equation}
\rhosfr({\rm \msun Myr^{-1} pc^{-3}}) = 0.12 \eff \rhogas^{1.5}(\msun {\rm pc^{-3})
}
\label{voltheoryeq}
\end{equation}
For comparison to data, we take $\eff = 0.01$ (Krumholz and McKee 2005).

The other model builds on the observations of density thresholds in 
nearby clouds (\citealt{2010ApJ...723.1019H,2010ApJ...724..687L})
to argue that the star formation rate is linearly proportional
to the mass of gas above a surface density or volume density threshold.
Based on studies of nearby clouds, \citet{2012ApJ...745..190L}
suggest the following relation:
\begin{equation}
\sfr(\msunyr) = 4.6\ee{-8} \fdense \mmol(\msun)
\end{equation}
or
\begin{equation}
\sfr(\msunmyr) = 0.046  \mdense(\msun)
\end{equation}
It would reproduce the relation with dense gas seen 
in high-mass star forming regions in the \mw\ (equation \ref{wueqn})
if the conversion
from far-infrared luminosity to star formation rate is scaled up by
a factor of 3.8.  Alternatively, the relation between dense gas as 
measured by extinction versus HCN \jj10\ luminosity could
be adjusted.

While these two models agree that star formation is strongly concentrated
in denser gas, they differ in detail. 
\citet{2012ApJ...745..190L} have argued
that their scaling law can be compatible with a volumetric law only if
$x = 1$ and $\rhogas > \rhoth$, where $\rhoth$ is a threshold volume
density that corresponds on average to the surface density threshold.

Other models have also been suggested, mostly involving timescales
different from the free-fall time. On a galactic scale, the orbital time
(\torb) could be relevant, and various studies have found good correlations
between \sigmasfr\ and \sigmagas /\torb\
\citep{1998ApJ...498..541K}.
This relation may also be useful for understanding starburst galaxies
(\citealt{2010ApJ...714L.118D, 2010MNRAS.407.2091G}),
but it cannot be extended to the level of molecular clouds in the \mw\
(e.g., \citealt{2012ApJ...745...69K}).

A possibly more relevant timescale for individual clouds or clumps
is the crossing time (\tcross), defined by the size of a relevant
dimension divided by an appropriate speed, which could be the thermal
sound speed or, more likely, the turbulent mean speed 
\citep{2000ApJ...530..277E}.

%% file: tests.tex

\section{Tests of Models}

\subsection{Uncertainties}

The uncertainties used in this section can be divided into two types,
which are discussed in much more detail in the Appendix. The first
type is an observational uncertainty which may vary from cloud to 
cloud or from region to region. These would affect the correlation
coefficients and the slope of linear fits, so they are included in 
figures and fits.
The second type of uncertainty is a systematic uncertainty, which
affects all the clouds and regions in the same way. These are not
included on the individual points in the plots nor in the fits, 
but they are shown separately in some plots.

\subsection{Tests of the free-fall model}

First, we test the pure form of the volumetric star formation model 
(equation \ref{voltheoryeq}) by
plotting \rhosfr\ versus \rhogas\ for the regions of increasing surface
density in nearby clouds. To do so, we convert the surface densities
of both gas and star formation rate into volume densities by dividing
by a length scale. Without further information on the depth of the cloud,
we assume a set of nested spherical shells with volumes defined by 
\begin{equation}
V(shell) = V_i - V_{i+1}
\end{equation}
with the index $i$ increasing with the extinction of the contour.
If there is only one YSO in the contour, we assume an uncertainty of
one YSO and plot only the point and an upper limit, corresponding to 
$1\pm1$ YSO. If there are no YSOs
in the contour, we set the number to unity and plot the point and an
upper limit based on $1\pm1$ YSOs. 
This approach allows us to show
upper limits on log-log plots, following \citet{2010ApJ...723.1019H}.

The results for Class I sources
are plotted in figure \ref{volsI}, along with the best fit relation and the
predictions from equation \ref{voltheoryeq} for an efficiency of 0.01.
We do not include the upper limits (for zero or 1 YSO) in the fit, but they
appear to be consistent.
For reference, the conversion to number density
(which includes all species) of particles is
\begin{equation}
n(\cmv) = 17.2 \rhogas(\msun {\rm pc^{-3}}).
\end{equation}
There is a strong correlation of \rhosfr\ and \rhogas, but it is 
steeper than predicted by the volumetric model. 
The best-fitting slope is
$2.02 \pm 0.07$, more than 7 $\sigma$  greater
than the value of 1.5 predicted by the free-fall picture. 
A fit using robust estimation (minimizing the absolute
deviation, without considering uncertainties), with the method
of \citet{1992nrfa.book.....P},  yields a slope of 1.95. 
The prediction of equation \ref{voltheoryeq} with $\eff = 0.01$ does get
the mean value about right, but does not predict the trend.

However, the assumption of spherical symmetry for
each shell could introduce errors and create an unrealistically
steep correlation. 
\citet{2012ApJ...745...69K} also noted the observational 
difficulties of comparing
volume densities, and they assumed that the volumetric law translated into
a surface density relation of the same form. In particular, they predicted
a linear relation between \sigmasfr\ and \sigmagas/\tff. 
This is easier to test as
what we directly measure is surface density of gas via extinction and
surface density of stars by counting in contours of extinction.
We do have to use volume density to compute the free fall time.
The result is shown in figure \ref{sdtI}, once again with the theory
plotted assuming $\eff = 0.01$. 
In this plot, the theory predicts
the mean star formation rates fairly well, but the best fit slope 
is $1.47\pm 0.06$, larger than the predicted slope of unity by 7
$\sigma$. Robust estimation yields 1.32.

\citet{2012ApJ...745...69K} actually used the properties of the whole
cloud rather than within contours of extinction (their table 2). 
In this case, all YSOs, not just the Class I objects, can be used,
because even Class II sources are unlikely to leave the entire cloud.
The free-fall time is computed from the mean
cloud density. We plot the same quantities as \citet{2012ApJ...745...69K},
but using the latest YSO catalogs, in Fig. \ref{cloudtffav2}.
As in Fig. \ref{sdtI}, the theory predicts the mean \sigmasfr\ reasonably correctly
but in this case, there is no convincing correlation in the data
(Pearson's $r = 0.35$; for this number of data points $r > 0.61$ is
required for a statistically significant correlation). 
This is a bit puzzling because the points indicated by ``MW clouds"
in Fig. 3 of \citet{2012ApJ...745...69K} do appear to be correlated.
This impression of a correlation arises because
\citet{2012ApJ...745...69K} plotted the same clouds twice, but
the second time with values taken from a contour with higher extinction 
values.  
Plotting the same clouds twice  in Fig. 3 of \citet{2012ApJ...745...69K}
effectively mixes our Fig. 1, which shows a very steep 
correlation {\it within} clouds, and Fig. 3, 
which shows no convincing correlation from one cloud to another.
\citet{2013ApJ...778..133L}
have also emphasized that there is a Schmidt relation (equation 1)
{\it within} clouds but not {\it between} clouds.

\subsection{Tests of a Crossing Time Model}

We have also tested the crossing time instead of \tff\ as the 
relevant timescale. We define \tcross\ as the time for a disturbance
traveling at the  turbulent
equivalent of the mean speed to cross the entire cloud (twice the size,
which is defined as the equivalent of the radius).
The turbulent speed was calculated from the linewidth measured for
the cloud, preferably an average over the cloud in the \coo\ \jj10 line,
but occasionally resorting to other tracers (see the Appendix for the
sources of velocity data). The mean speed, related to the
FWHM linewidth by a factor of 0.678, was used for the calculation.
The plot of log \sigmasfr\ versus log( \sigmagas/\tcross) 
(Fig. \ref{tcross}) shows no correlation at all, with Pearson's $r = 0.22$.
Crossing times may be relevant on larger or smaller scales, but we find no
evidence that they matter on the scales of the nearby clouds.

\subsection{Tests of Threshold Idea}

For the clouds as a whole, \citet{2010ApJ...724..687L} have
already provided the evidence for a better correlation between
star formation and cloud mass if the cloud mass is restricted to
the mass above a threshold surface density. Here we look at the
data in other ways. 

We divide each cloud into two regions separated by the contour of
$\av = 8$.
Both
\citet{2010ApJ...724..687L}
and 
\citet{2010ApJ...723.1019H}
have suggested a threshold near this value of extinction.
For $\av < 8$, we will use the term ``low" and for $\av \ge 8$, we
will use the term ``high."
For comparison to large-scale studies, we aggregate all the YSOs and mass 
into the sum over each region. We do the same for Class I sources alone.
We then compute a single value for the surface density of the star
formation rate by 
dividing the total, based on the total number of YSOs, a mean
mass of 0.5 \msun, and a duration of the infrared excess of 2 Myr
(0.55 Myr for the Class I sources), and assign uncertainties in these
mean values, based on counting statistics. We calculate the mean
surface density as the total mass divided by the total surface area
and assign an uncertainty as the range of mean surface densities 
for both low and high surface density regions.

The results (Fig. \ref{thresh}) are quite
striking. The mean surface density of YSOs in the high region is 6.7 times
the mean in the low region. For Class I sources, the ratio is 14.
These results are consistent with the proposal of a threshold
at $\av \sim 8$ mag.
In terms of raw numbers, of the 2915 YSOs in the new catalogs, 64\% are 
projected onto high extinction parts of the cloud.
Since some of the
older YSOs could have left their birthplaces, this is a lower limit.
If we restrict attention to Class I sources, the fraction rises to 
77\%. The cumulative area of all the clouds above this surface
density is only 20\% of that in all the clouds above $\av = 2$, 
and the cumulative mass is only 38\% of the total above $\av = 2$.
The result is further diluted by the fact that two large clouds were not
mapped by Spitzer down to $\av = 2$, so the total area and mass below
$\av = 8$ is undercounted. Finally, there is probably at least a factor
of two more mass below $\av = 2$ in CO-emitting gas 
\citep{2008ApJ...680..428G}.
In summary, the great majority of star formation occurs in a small fraction
of the cloud area or mass.

The most direct test of the \citet{2012ApJ...745..190L} proposal
(equation 8 or 9) is to plot the star formation rate versus
the mass of dense gas. 
However, a plot of SFR versus gas mass naturally shows a correlation
because big clouds tend to produce more stars.
Instead we first plot the ``efficiency" in the sense of SFR in the dense
gas over the mass of dense gas (\sfr / \mdense) versus the mass of dense gas 
(plotting logarithms) and look at the scatter (Fig. \ref{leeidea}). 
We used the total number of YSOs within the
$\av = 8$ mag contour to measure the SFR and the mass inside that
contour to measure the mass of dense gas. 
The mean and
standard deviation of log(\sfr /\mdense ) is $-1.61\pm0.23$. The dispersion is 
 comparable to the observational uncertainties 
and the likely systematic uncertainties (about 0.3 in the log). 
In contrast, if we plot the same quantities for the total
SFR in the whole cloud and the total mass of the cloud, we obtain
$\mean{{\rm log} \sfr/ \mcloud} = -2.06\pm 0.83$, resulting in a standard
deviation that is 3.6 times larger 
and substantially larger than both observational and likely 
systematic uncertainties (Fig. \ref{leecloud}). 
The dense gas clearly yields a much more accurate prediction of the SFR.
No obvious trend in the efficiency is seen in either figure \ref{leeidea}
or \ref{leecloud} over more than two orders of magnitude in the 
corresponding gas mass, consistent with a roughly linear proportionality.
Versions of these figures with data from additional clouds studied by 
\citet{2010ApJ...724..687L}
added are shown in 
\citet{2013arXiv1312.5365P}.
While further
assumptions are needed to add those clouds, they are consistent with
the clouds shown here.

Having shown that the dispersion of star formation rate over dense
gas mass is quite low, we now plot directly the star formation
rate versus the mass of dense gas (equation 9).
The results (black points) are plotted
in Figure \ref{ladatesthigh}. The points show a strong correlation
and lie reasonably close to the line representing equation 9.
Fitting the data, we obtain 
$ \sfr (\msunmyr) = 0.041 \mdense^{0.89}$ with a correlation coefficient
of 0.963. The coeficient is 0.9 times that found by 
\citet{2012ApJ...745..190L}, using a somewhat different set of
clouds and threshold.

\citet{2012ApJ...745..190L} pointed out that their equation has a
coefficient about 3.8 times that found by \citet{2005ApJ...635L.173W}.
\citet{2013ApJ...765..129V} have reexamined some of the massive dense
clumps studied by \citet{2005ApJ...635L.173W} to get better SFR by
using radio continuum and integrating over the whole \ion{H}{2} region.
We plot these SFRs in red using the mass from the maps of HCN \jj10\
from \citet{2010ApJS..188..313W}, where both measurements exist.
These are broadly consistent with the prediction now, but there is 
a caveat. Massive stars destroy their environment rapidly, so the
\ion{H}{2}\ region is substantially larger than the remaining
dense clump. Thus the original mass of dense gas was probably larger,
which would move the massive dense clumps to the right in the figure.
If we ignore that issue and fit the combination of the nearby clouds
and the massive clumps, the result is
$ \sfr(\msunmyr) = 0.040 \mdense^{0.90}$ with a correlation coefficient
of 0.924. The normalizing coefficient is close to that
of  \citet{2012ApJ...745..190L}.

%% file: disc.tex

\section{Discussion} \label{disc}

The tests presented above do not favor a picture in which the
free fall time, computed from the mean density of a cloud, 
is an important factor in predicting the star formation rate within
the cloud.  This result
is not surprising because clouds do not appear
to be collapsing at free fall (e.g., \citealt{1974ApJ...192L.149Z}). 
When the free-fall time is computed
more locally, in nested contours, and a low efficiency factor per
free-fall time
is included, models that predict $\rhosfr \propto  \rhogas^{1.5}$ get
the magnitude of $\rhosfr$ roughly right, but the actual $\rhosfr$
increases more rapidly ($\rhosfr \propto  \rhogas^{2}$). 
The free-fall time may become
relevant, if ever, only on scales of individual infalling envelopes.
The cloud crossing time also does not seem to be very useful as
a predictor of star formation rates, though crossing times may be
relevant on larger or smaller scales.

The threshold model is really just a codification of observational facts.
Star formation is observed to be highly concentrated in regions of
high surface density, more concentrated than can be explained with
models using a free-fall time calculated from a mean density.
It is important to clarify what 
\citet{2012ApJ...745..190L} meant by a threshold. They say
``Furthermore, our data are consistent with the existence of a
column density threshold for star formation activity above which
the SFR appears to be linearly correlated with the total cloud mass
above the threshold."
They did not claim that there was {\bf no} star formation below 
the threshold. 
Similarly 
\citet{2010ApJ...723.1019H} said 
``A steep increase and possible leveling off in \sigmasfr\ at a
threshold $\Sigma_{thresh} \sim\ 100$ to 200 \msunpc\ is seen..."
One might interpret these statements as advocating
a step function or a ``precise" boundary between star-forming
and non-star-forming gas. That was not the intent of either set of authors.

Indeed, \citet{2013ApJ...778..133L}
have tested models with a step function
(Heaviside function) in 4 clouds, finding that two are consistent with
such an extreme definition of threshold, while two are not. 
They conclude
that a Heaviside function, while possible in some clouds, is too extreme
for a general definition of threshold.

A second point that needs clarification is the scale over which
the gas surface density is measured. One could argue that a surface
density threshold is a tautology because stars can only form in 
dense gas. The scale over which the surface density is measured
in the nearby clouds is limited by the resolution of the extinction maps
to 270\arcsec, which corresponds to 0.34 pc at the mean distance.
The mean area of the $\av = 8$ contour corresponds to a radius of 2.5 pc.
The gas density measured from extinction corresponds to scales of clumps
which may contain many individual dense cores, the sites of individual
star formation. There is no a priori
reason why such dense cores cannot form in regions of lower average
surface density, in which case they could be spread more evenly over
the cloud. Some occasionally do form in regions of lower extinction,
but they are rare.
To be specific, 
\citet{2008ApJ...684.1240E} found that 75\% of prestellar cores were found
above $\av = 6.5$, 8, and 19.5 mag in Perseus, Serpens, and Ophiuchus,
respectively. No cores were found below $\av = 7$ or 15 mag in Serpens and
Ophiuchus, respectively. The constraints on cores hosting protostars were
even tighter.

The issue of scale is also relevant to the interpretation of
the continued rise of star formation rate surface density roughly
proportional to the square of gas surface density, even above the
threshold of $\av = 8$ mag, in young clusters
\citep{2011ApJ...739...84G}.
Our data do not disagree with this result, but we consider it to be
probing the distribution of star formation {\it within} a clump rather than 
constraining the definition of a star forming clump, defined on larger scales
to include the bulk of star formation.

One might also question whether the star formation rates in these
clouds are low because they are young, and star formation accelerates.
While observers lack the ability to determine cloud ages, we can make
a crude measure by examining the class distribution of the YSOs.
For the clouds as a whole, we computed an ``age indicator" from
the relative number of sources in older and younger classes as follows:
\begin{equation}
Age = [N(YSOs)-N(I)-N(F)]/N(YSOs)
\end{equation}
where $N(YSOs)$ is total number of YSOs, $N(I)$ is the number of Class I
objects, and $N(F)$ is the number of Flat spectrum sources. All but 4 clouds
have ``age" greater than 0.5, indicating a preponderance of Class II and Class
III objects, and hence that star formation has been proceeding for at least
1-2 Myr. Three of the four exceptions are small clouds with only one or two
YSOs; the fourth, with ``age" of 0.5, is IC5146 NW, with 38 YSOs.
For clouds with a sufficient number of YSOs to determine the ``age",
there is no obvious evidence for acceleration, but the dynamic range is
too small to provide a strong test.

While a complete survey of theoretical ideas is beyond the scope of this
paper, a few can be mentioned that address the observations discussed
here.
\citet{2013ApJ...773...48B} have argued that a model of formation
of molecular clouds from galactic hydrodynamical flows and subsequent
gravitation collapse, while dense gas continues to form during star
formation, can match the observations presented by 
\citet{2010ApJ...723.1019H}. 
\citet{2012ApJ...761..156F} show that models and simulations with MHD
turbulence and local free-fall times also match the data, with certain
assumptions. Our results suggest that the free-fall time calculated from
large-scale mean densities is not a useful parameter for predicting
star formation rates, but \tff\ calculated more locally may be relevant,
as suggested by 
\citet{2011ApJ...743L..29H}. 
\citet{2012ApJ...761..156F} also focus on the
remaining scatter in the \citet{2010ApJ...723.1019H} data, which
is indeed substantial. They suggest that virial parameter, Mach number,
the nature of turbulent forcing, and magnetic field all contribute to
this scatter. At this point, our knowledge of these parameters for
these clouds does now allow sharper tests of that picture. In particular,
the small variation in linewidth among this sample (Table \ref{tbl-1})
provides too small a lever arm. However, we can say that testing
the dependence on these parameters will be better done after the first
order dependence on cloud, or better dense gas, mass has been removed.

Theorists modeling gravoturbulent fragmentation
generally argue that volume density, rather than column density,
should be the controlling factor (see 
\citealt{2013arXiv1312.5365P}
for a review), 
and many theories invoke a critical volume density for star formation.
\citet{2012ApJ...745..190L}
suggest a volume density of $n \sim \eten4$ \cmv\ as the value that 
corresponds to the surface density  threshold.
The average volume density in the regions above the $\av = 8$ mag threshold
in this sample is $\mean{n} = (6.1\pm4.4)\ee3$ \cmv, roughly consistent, but
lower than \eten4 \cmv\ and with a rather large dispersion. 
At this point, column density, rather
than volume density, looks more likely to be the controlling factor.
As discussed in more detail by 
\citet{2010ApJ...723.1019H},
models in which magnetic fields
\citep{1976ApJ...210..326M}
or photoionization
\citep{1989ApJ...345..782M}
regulate star formation, while out of theoretical fashion, are still
the models most consistent with the actual observations.
A more recent treatment also shows the relevance of column density
via self-shielding, and offers a suggested explanation of the threshold
value around $\av = 8$ mag
\citep{2013arXiv1306.5714C}.

While the structures bearing prestellar or protostellar cores often
appeared filamentary in the extinction maps, the resolution was insufficient
to delinate the structure well. More recent mapping with Herschel shows that 
prestellar and protostellar cores are closely confined to narrow 
filaments with surface density corresponding to $\av = 10$ mag
\citep{2010A&A...518L.102A}
and average width (FWHM) of $0.10\pm0.03$ pc
\citep{2011A&A...529L...6A}.
In contrast, filaments with $\av \sim 2$ mag and similar or smaller width
contained no star formation
\citep{2010A&A...518L.102A}.
Further study of these filaments is likely to lead to a deeper understanding
of the threshold picture
(e.g., \citealt{2013A&A...553A.119A}).
Testing for a threshold in mass per unit length along a filament will be
necessary.

%% file: conclusions.tex

\section{Conclusions} \label{conclusions}

While a star formation relation employing a mean \tff\ seems
to work on very large scales of galaxies \citep{2012ApJ...745...69K}
and to predict roughly the mean rate of star formation in nearby
clouds, with $\eff = 0.01$ \citep{2005ApJ...630..250K}, it does 
not predict the behavior of star formation rates on small scales
within molecular clouds. These observed rates show a steeper dependence
on the surface density or volume density per free-fall time than predicted.
When applied to whole clouds, the data show no significant correlation
with \sigmagas/\tff, with  \tff\ calculated from the mean cloud
density, as advocated by \citet{2012ApJ...745...69K}. 
If the crossing time is substituted for the free fall time, no correlation
at all is seen.

In contrast, the picture in which the star formation rate is linearly
proportional to the mass of dense gas above a threshold surface density
\citep{2012ApJ...745..190L},
corresponding to $\av = 8$ mag, matches the data for nearby clouds with
very low dispersion.
Using new determinations of the star formation rate in massive dense clumps
\citep{2013ApJ...765..129V}, those data are matched reasonably well also
by the same relation (equation 9), 
but there are more
uncertainies in those distant clumps, where counting of YSOs is not
practical. Since this same relation works well for galaxies, using
the HCN emission to estimate crudely the mass of dense gas, it would
appear to be the best relation to use in simulations of sufficient
resolution to locate the dense gas.
However, the particular threshold that applies to nearby clouds may not apply
in other regions, especially those with lower metallicity or much stronger
radiation fields (cf. clouds in the CMZ, \citealt{2013MNRAS.429..987L}).
In any case, removing the first order dependence on the mass of dense gas
is necessary before testing the importance of other factors suggested
by theorists.

%% file: ack.tex

We thank L. Hartmann, P. Padoan, and C. Federrath for helpful
discussions and M. Dunham for compiling the YSO lists for the
Gould Belt project.
This research was supported by 
NSF Grant AST-1109116 to the University of Texas at Austin.
NJE thanks the European Southern Observatory, Santiago,
for hospitality during an extended visit when this work was begun.

%% file: appendix.tex
\appendix{}



\section{Appendix A: Information on Clouds}

The two pieces of information about the clouds from outside the
c2d or Gould Belt data are the distances and velocity dispersions.
The distances are taken from \citet{2013AJ....145...94D}, who
give references, but we also give
separate distances to the five clouds in Cepheus, based on
\citet{2009ApJS..185..198K}.

The velocities in Table \ref{tbl-1} are the full width at half
maximum of the \coo\ \jj10\ line, whenever possible. Ideally, these are
based on an average over the whole cloud, but in many cases, only
the linewidths for individual points are available.  In that
case, the values in the table are averaged by hand.
In some cases, the linewidths are at best educated guesses,
based on other isotopologues or species.
Brief discussions of each cloud with references are given below.

\subsection{Aquila}
This region lies in the ``Serpens-Aquila rift" and has been called
``Serpens South", ``Serpens-Aquila", and ``Aquila." We adopt the
last convention.
\citet{2008ApJ...673L.151G} describe the Spitzer data, and 
\citet{2011A&A...535A..77M} present millimeter continuum data.
The distance is taken from \citet{2011A&A...535A..77M},
which places it closer to us than the Serpens cloud, formerly
considered to be at the same distance.
No data on \coo\ \jj10\ were found, so a linewidth of 3.0 \kms\
was estimated from line profiles of \hcop\ \jj43\ found in 
\citet{2011ApJ...737...56N}. This estimate is clearly quite
uncertain.

\subsection{Auriga}

This region was identified as a single entity composed of a number
of Lynds dark clouds by 
\citet{2009ApJ...703...52L}.
It has also  been called the ``California Cloud", or the
``California-Auriga" cloud \citep{2013ApJ...764..133H}.
Some information on  \coo\ found for this cloud refers to
NGC5179, the nebula around Lk H$\alpha$ 101, where \dv (\coo)
was measured to be about 2 \kms\ based on two spectra in
\citet{1976ApJ...206..443K}.
These may not be representative of the rest of the cloud.
A different part of the cloud was mapped in \coo\ by 
\citet{1991A&A...249..483H},
including the region around L1442, L1449, and L1456.
Based on averaging their Gaussian standard deviations and multiplying
by 2.35 to get FWHM, values of $\dv = 1.41$ to 1.69 \kms\ were obtained
for parts of the cloud. We adopt 1.7 \kms, based on these fragmentary
measurements, but we emphasize the uncertainty.

\subsection{Cepheus clouds}

For the Cepheus flare, 
\citet{2009ApJS..185..198K} has published analysis of the
six separate regions, including separate distance and velocity
measurements. The clouds are labeled by Lynds numbers. We have
made the following associations with our Spitzer regions:
Ceph-1 corresponds to L1251$+$L1247 with a linewidth of \coo\
of 1.9 \kms; Ceph-2 corresponds to L1241 with \dv (\coo) of 2.2 \kms;
Ceph-3 corresponds to  L1172$+$L1144 with \dv (\coo) of  1.6 \kms;
Ceph-4 corresponds to  L1148$+$L1152+L1155 with \dv (\coo) of 1.0 \kms;
Ceph-5 corresponds to L1228 with \dv (\coo) of 1.6 \kms.

\subsection{Chamaeleon}

\citet{1994ApJ...433...96V} give mean \dv (\coo) for Cha II and
III, but not for Cha I.  Cha I was mapped by 
\citet{1998ApJ...507L..83M} in \coo, but no information on linewidth
was given.  It was also mapped in CO by 
\citet{2001PASJ...53.1071M}, who give a \dv\ for CO of 2.0 \kms. 
They also give 2.8 \kms\ and 2.6 \kms\ for Cha II and III respectively. 
Based on ratios of the \coo\ to CO line widths in those clouds of 0.426, 
we used $0.426\times 2.0 = 0.85$ \kms\ for \coo\  in Cha I.

\subsection{Corona Australis}

The cloud has been studied in considerable detail, but we had
to go back to
\citet{1979ApJ...227..832L} to get data on \coo.
From a map of \dv\ of \coo\ \jj10\  over the cloud, one can see
values from 1.0 to 2.0 \kms, with 1.5 \kms\ being the most 
characteristic value. 

\subsection{IC5146}
The velocities are from \citet{1992AJ....104.1525D}.
For IC5146 E, we use cloud E, which has  $\dv = 2.45\pm 0.3$;
for  IC5146 NW, we use cloud C, which has $\dv =  2.04\pm0.42$,
where these are the average and standard deviation of all relevant
\dv (\coo) in their Table 1.

\subsection{Lupus}
\citet{1999PASJ...51..895H} mapped the area in \coo\ \jj10, obtaining
typical \dv\ of 1.2 \kms, but there is considerable variation in the region. 
We now adopt values for each cloud separately. We use
\citet{1999PASJ...51..895H} for Lupus V and VI.
For Lupus I, III, and IV, the linewidths come from \coo\ \jj21\
maps by 
\citet{2009ApJS..185...98T}, but averaging the mean and median
linewidths (N. Tothill, private communication) as was also done
for Ophiuchus, Perseus, and Serpens (below).

\subsection{Musca}
\citet{1994ApJ...433...96V} gives \dv (\coo) of 0.8 \kms\ for Musca.

\subsection{Ophiuchus}

The linewidth information was given in
\citet{2009ApJS..181..321E}, originally provided by J. Pineda
(2008, private communiction) using the COMPLETE maps 
\citep{2006AJ....131.2921R}.
The values represent the average of the mean and median linewidths
averaged over the clouds.

\subsection{Ophiuchus North}

This region was identified as Scorpius in the Gould Belt nomenclature,
but it is now known as Ophiuchus-North, since it is not actually in Scorpius.
\citet{2012ApJ...754..104H} published the Gould Belt data under this name.
This region was mapped by 
\citet{1991ApJS...77..647N} in \coo, and the identification of cores
in the notation of \citet{1991ApJS...77..647N} with the Spitzer
regions was made using the table in \citet{2012ApJ...754..104H}.

\subsection{Perseus}

The linewidth information was obtained in the same way as for
Ophiuchus.

\subsection{Serpens}

The linewidth information was obtained in the same way as for
Ophiuchus. 
The distance is updated from \citet{2010ApJ...723.1019H}
based on the VLBA parallax of the
Herbig Ae/Be star, EC95, yielding a distance of $429\pm2$ pc
\citep{2011RMxAC..40..231D}.

\input errors.tex

%% file: errors.tex


\section{Appendix B: Error Propagation}

The uncertainties plotted on each point in a plot are based on observational
uncertainties only. In some plots, a separate uncertainty is plotted to
indicate the estimated systematic uncertainties for the sample as a whole.
The basic observational quantities are the number
counts of YSOs, the extinction, and the distance. The
uncertainty in the extinction propagates immediately into the
mass surface density of gas, while the distance uncertainty
propagates into the uncertainty in the area, 
through the conversion of angles to linear area.
The further propagation
of observational uncertainty in those quantities 
into the derived quantities is explained
next, followed by a discussion of systematic uncertainties.

For the star formation rate, the uncertainties are
based on counting statistics:
\begin{equation}
\sigma(\sfr) = (\sqrt{N}/N) \sfr
\end{equation}
where $N$ is the number of YSOs in that sample (Class I for Figs. 1 and 2
and the upper panel of Fig. 5; all classes for the other figures).

For $\sigmasfr = \sfr/A$, where $A$ is the area, the uncertainties include
the uncertainty in the distance via the area: 
\begin{equation}
\sigma(\sigmasfr) = \sigmasfr
\left[\left(\sigma(\sfr)/\sfr\right)^2 + \left(\sigma(A)/A\right)^2\right]^{0.5}
\end{equation}

For $\rhosfr$ (Fig. 1), we express it as
\begin{equation}
\rhosfr = \sfr/V = \frac{\sfr}{0.752 A^{1.5}}
\end{equation}
where $V$ is the volume. Then
\begin{equation}
\sigma(\rhosfr) = \rhosfr \left[\left(\frac{\sigma(\sfr)}{\sfr}\right)^2
+ \left(\frac{1.5 \sigma(A)}{A}\right)^2\right]^{0.5}
\end{equation}

For masses (\mcloud\ or \mdense ), the uncertainties include the uncertainty
in the extinction from the extinction maps and the distance uncertainty through
the uncertainty in the area since mass is measured from extinction times area.
Observational uncertainties were not available for the massive dense
cores plotted in figure 8, so we assumed uncertainties of 0.30 in 
log(\sfr) and 0.15 in log(\mdense).

For surface densities of gas, the distance uncertainty does not enter because
both mass and area scale as $d^2$, so the uncertainty is just the uncertainty 
in the extinction maps. 

For volume densities of gas (Fig. 1), the distance uncertainties do
enter. We express the gas density as
\begin{equation}
\rhogas = M/V = \frac{\sigmagas}{0.752 A^{0.5}}
\end{equation}
In this form,
\begin{equation}
\sigma(\rhogas) = \rhogas \left[\left(\frac{\sigma(\sigmagas)}{\sigmagas}\right)^2
+ \left(\frac{0.5 \sigma(A)}{A}\right)^2\right]^{0.5}
\end{equation}

For figure 3, where we plot $\sigmagas/\tff$,
we can write
\begin{equation}
\sigmagas/\tff = \frac{\sigmagas \rho^{0.5}_{\rm gas}}{8.08} = 
0.143 \sigmagas^{1.5} A^{-0.25}
\end{equation}
using
\begin{equation}
\rho_{\rm gas} = \frac{M_{\rm gas}}{0.752 A^{1.5}} 
= \frac{\sigmagas}{0.752 A^{0.5}}
\end{equation}
Then
\begin{equation}
\sigma(\sigmagas/\tff) = \frac{\sigmagas}{\tff}
 \left[\left(\frac{1.5\sigma(\sigmagas)}{\sigmagas}\right)^2 + 
\left(\frac{0.25\sigma(A)}{A}\right)^2\right]^{0.5}
\end{equation}
so, the distance uncertainty enters weakly, only through the area.

For figure 4, the distance enters via the size, which depends
on the square root of the area. Using the facts that the mean
speed is $0.678 \dv$ and that 
1 \kms\ is 1.023 pc Myr$^{-1}$,
we can write
\begin{equation}
\sigmagas/\tcross = \frac{\sigmagas \mean{v}}{2 r}
= 0.6165 \frac{\sigmagas \dv ({\rm \kms})}{A^{0.5}}
\end{equation}
and 
\begin{eqnarray}
\sigma(\sigmagas/\tcross) = \frac{\sigmagas}{\tcross} \times \qquad \quad
 \nonumber \\
 \left[\left(\frac{\sigma(\sigmagas)}{\sigmagas}\right)^2 + 
\left(\frac{0.5\sigma(A)}{A}\right)^2
+ \left(\frac{\sigma(\dv)}{\dv}\right)^2\right]^{0.5}.
\end{eqnarray}
\\

The dominant source of uncertainty is the $\dv$, and most lack
reliable uncertainties. We have taken an uncertainty of 30\% for each
cloud in $\sigma(\dv)$.

The uncertainties in quantities were propagated to asymmetric
uncertainties in the
logarithms for the plots by taking the logarithms of the minimum
and maximum values of the quantity. Because the fitting routine could
use only symmetric uncertainties, we used the maximum of the two
asymmetric errors for the fit.

In addition to the observational uncertainties, there are systematic
uncertainties that affect all points in the same way. These are not
included in the error bars on each point because they do not affect the
issue of correlations or slopes. They would affect scaling of the axes,
offsets for fits, and the absolute value of the mean values in figures 6 and 7.

The star formation rate is computed from the counts of YSOs, a mean mass per
star, and a timescale over which those YSOs are 
visible \citep{2009ApJS..181..321E}.
As noted earlier, the counts of YSOs could be low by about 30\%
if less cautious  methods of removing contaminants are adopted. 
The counts of Class I objects could decrease by about 50\% if many sources
classified as Class I are found not to be associated with dense gas, but in that case
the timescale would be decreased by the same factor, leaving the rate
unchanged.
The mean stellar mass is taken to be 0.5 \msun, and the half-life for YSOs,
as identified by the c2d and Gould Belt programs, was taken to be 2 Myr.
The half-lives of earlier stages are scaled to that value, and we used 0.55 Myr
for Class I objects. The mean stellar mass is taken from IMF models.
The uncertainties in these numbers are probably a factor of 2, which propagates 
directly into \sfr\ and \sigmasfr.

For the masses of gas, the conversion of extinction to surface density or mass
is the main systematic uncertainty. We use the Case A models of Weingartner
and Draine for $R_V = 5.5$, as updated on the website noted in \S \ref{intro}. 
These are consistent with our earlier paper \citep{2010ApJ...723.1019H}, 
and more or less consistent with \citet{2010ApJ...724..687L}. However, in 
a more recent paper, 
\citet{2013ApJ...778..133L}
use a different model of dust, closer to Case B models. 
These match data well \citep{2013A&A...549A.135A},
but have some theoretical difficulties (Draine, personal communication). The Case B
models imply a higher ratio of $\sigmagas/\av$ by a factor of 1.49. This is not
a two sided error, but instead a possible increase in  \mcloud,
\mdense, and \sigmagas, all by the factor 1.49. The effect on
$\sigmagas/\tff$ would be an increase by a factor of $1.49^{1.5} = 1.82$.

%% file: table1.tex
\begin{deluxetable}{lrrrrrrrcrl}
\tabletypesize{\scriptsize}
\tablecaption{Basic Data on Clouds\tablenotemark{1}
\label{tbl-1}}
\tablewidth{0pt}
\tablehead{
 \colhead{Cloud} & \colhead{Dist.} & \colhead{Size} & \colhead{\sfr} &
 \colhead{\mcloud} & \colhead{\mdense} & \colhead{\sigmasfr}
 & \colhead{\sigmagas} & \colhead{\tff} & \colhead{$\Delta$v} & 
 \colhead{\tcross} \\
 \colhead{-}  & \colhead{(pc)} & \colhead{(pc)}  &  \colhead{(\msun} &
\colhead{(\msun)}  & \colhead{(\msun)} &
\colhead{(\msunmyr } & \colhead{(\msun} & \colhead{(Myr)} & 
\colhead{(\kms)} & \colhead{(Myr)} \\
 \colhead{ } & \colhead{ } & \colhead{ } & \colhead{ Myr$^{-1}$) } &
  \colhead{ } & \colhead{ } & \colhead{pc$^{-2}$)} & \colhead{pc$^{-2}$)} &
  \colhead{ } & \colhead{ } & \colhead{ } \\
 }
\startdata
Aquila	    & 260	& 7.56	& 322.3	& 24446 & 16034 & 1.80	& 136.2	& 2.20	& 3.00	& 7.26 \\
Auriga N    & 450	& 1.31	& 0.50	& 503   & 13    & 0.09	& 92.78	& 1.11	& 1.70	& 2.22 \\
Auriga	    & 450	& 5.98	& 36.0	& 10391 & 1134  & 0.32	& 92.37	& 2.38	& 1.70	& 10.1 \\
Cepheus 1   & 300	& 1.75	& 8.50	& 671	& 5.5   & 0.89	& 69.91	& 1.48	& 1.90	& 2.65 \\
Cepheus 2   & 300	& 1.46	& 0.00	& 499   & 12    & 0.00	& 74.18	& 1.31	& 2.20	& 1.92 \\
Cepheus 3   & 288	& 1.75	& 10.5	& 633   & 41	& 1.09	& 65.49	& 1.53	& 1.60	& 3.16 \\
Cepheus 4   & 325	& 1.08	& 0.50	& 267   & 0	& 0.14	& 73.31	& 1.13	& 1.00	& 3.11 \\
Cepheus 5   & 200	& 1.07	& 4.75	& 233   & 31    & 1.33	& 65.32	& 1.19	& 1.60	& 1.92 \\
Cha I	    & 150	& 1.30	& 20.5	& 482   & 176	& 3.88	& 91.18	& 1.11	& 0.85	& 4.40 \\
Cha II	    & 178	& 1.78	& 6.00	& 637	& 64    & 0.61	& 64.34	& 1.55	& 1.20	& 4.26 \\
Cha III	    & 150	& 2.24	& 1.00	& 746   & 0	& 0.06	& 47.37	& 2.03	& 1.10	& 5.86 \\
Corona Aus. & 130	& 0.98	& 10.5	& 279   & 139	& 3.47	& 92.16	& 0.96	& 1.50	& 1.89 \\
IC5146 E    & 950	& 4.42	& 23.25	& 3365  & 0	& 0.38	& 54.79	& 2.65	& 2.45	& 5.20 \\
IC5146 NW   & 950	& 5.28	& 9.50	& 5179  & 92	& 0.11	& 59.13	& 2.79	& 2.04	& 7.46 \\
Lupus I	    & 150	& 1.68	& 3.25	& 512	& 39    & 0.37	& 57.86	& 1.59	& 2.17	& 2.23 \\
Lupus III   & 200	& 2.22	& 17.0	& 912   & 96	& 1.10	& 59.08	& 1.81	& 2.11	& 3.03 \\
Lupus IV    & 150	& 0.90	& 3.00	& 189	& 50    & 1.19	& 75.14	& 1.02	& 1.53	& 1.69 \\
Lupus V	    & 150	& 1.93	& 10.5	& 704   & 0	& 0.90	& 60.50	& 1.66	& 1.20	& 4.62 \\
Lupus VI    & 150	& 1.46	& 11.0	& 454   & 4.3	& 1.63	& 67.50	& 1.37	& 1.20	& 3.52 \\
Musca	    & 160	& 1.47	& 3.00	& 335   & 0	& 0.44	& 49.15	& 1.62	& 0.80	& 5.31 \\
Ophiuchus   & 125	& 3.08	& 72.75	& 3128  & 1209  & 2.44	& 104.8	& 1.60	& 0.94	& 9.45 \\
Oph North 1 & 130	& 0.49	& 0.25	& 66    & 11	& 0.33	& 86.97	& 0.70	& 0.80	& 1.77 \\
Oph North 2 & 130	& 0.46	& 0.00	& 68    & 26	& 0.00	& 102.7	& 0.63	& 1.37	& 0.97 \\
Oph North 3 & 130	& 0.94	& 1.25	& 258   & 64	& 0.45	& 94.13	& 0.93	& 1.00	& 2.72 \\
Oph North 4 & 130	& 0.58	& 0.00	& 76    & 0	& 0.00	& 71.83	& 0.84	& 1.23	& 1.37 \\
Oph North 5 & 130	& 0.39	& 0.00	& 34    & 0	& 0.00	& 69.65	& 0.70	& 1.23	& 0.93 \\
Oph North 6 & 130	& 0.70	& 0.75	& 116   & 12	& 0.48	& 74.85	& 0.90	& 0.85	& 2.38 \\
Perseus	    & 250	& 4.83	& 96.25	& 6586  & 2147  & 1.32	& 89.99	& 2.16	& 1.54	& 9.03 \\
Serpens	    & 429	& 3.92	& 56.0	& 6520  & 4213  & 1.16	& 135.1	& 1.59	& 2.16	& 5.23 \\
\hline
Median 	    & 150	& 1.47  & 6.0   & 504   & 39    & 0.48	& 73.3  & 1.48  & 1.50 	& 3.11 \\
Mean	    & 256	& 2.17	& 25.1	& 2355	& 885	& 0.89	& 78.6	& 1.47	& 1.52	& 3.99 \\
Stdev	    & 220   	& 1.83	& 61.7	& 4940	& 3044	& 0.99	& 22.2	& 0.58	& 0.56	& 2.59 \\
\enddata

\tablenotetext{1}{ These cloud values refer to extinction contours of $\av = 2$ (
$\av = 6$ for Serpens and $\av = 3$ for Ophiuchus) 
}
\end{deluxetable}

%% file: figs.tex

\clearpage

\begin{figure}
\center
\includegraphics[scale=0.5, angle=-90]{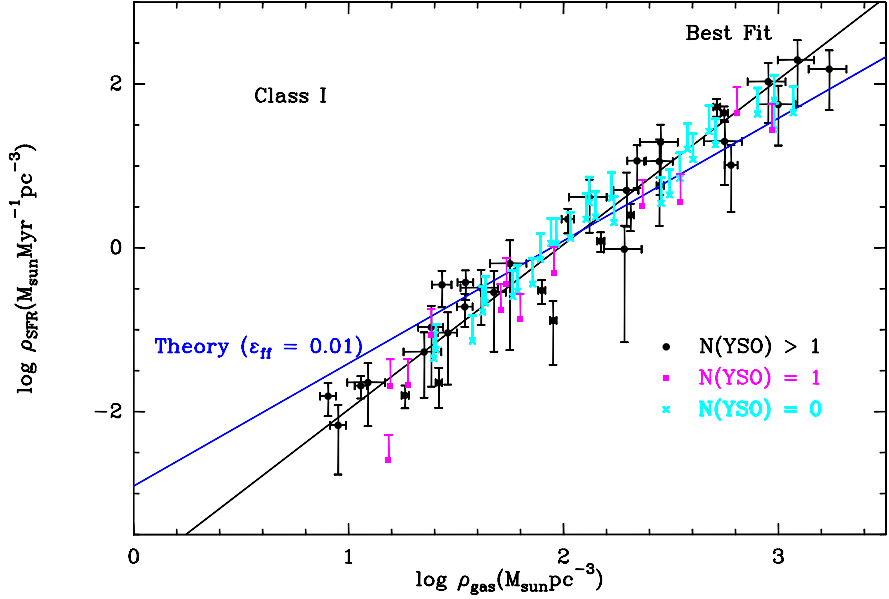}
\caption{
Plot of log(\rhosfr) versus log(\rhogas) for Class I sources in the
nearby clouds. 
The uncertainties are propagated into both axes for
contours with more than one YSO. For contours with only 1 YSO, only
upper limits are plotted since the uncertainties exceed the value.
Likewise, contours with no YSOs are plotted as if they had one YSO,
again with only upward uncertainties, assuming an uncertainty of 1
in the number of YSOs.
The blue line is the prediction of equation \ref{voltheoryeq}.
The black line is the result of a least-squares fit to the data.
}
\label{volsI}
\end{figure}

\begin{figure}
\center
\includegraphics[scale=0.5, angle=-90]{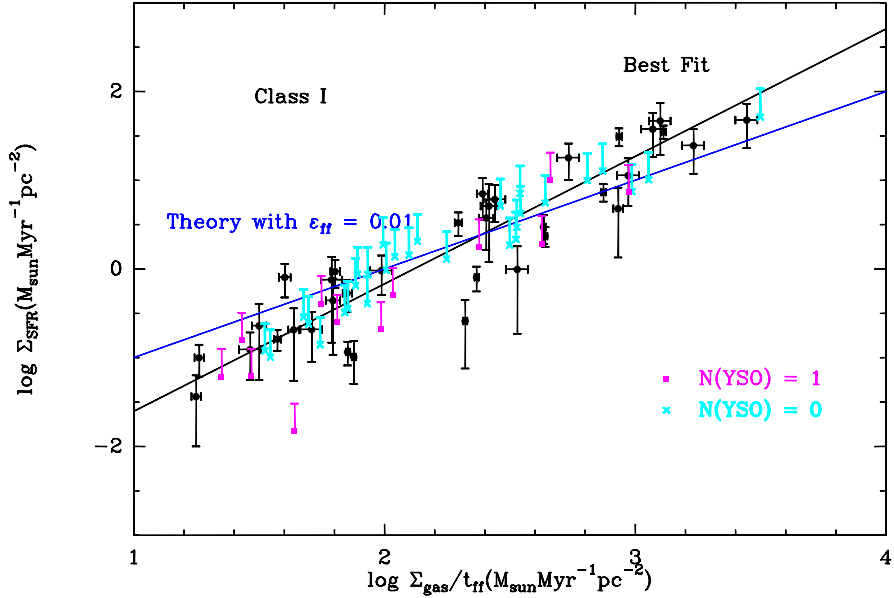}
\caption{
Plot of log(\sigmasfr) versus log(\sigmagas/\tff) for Class I sources in the
nearby clouds.  The uncertainties are propagated into both axes for
contours where the value exceeds the uncertainty. 
Only the points and the upper limits are plotted when the 
uncertainties exceed the value, but the data are consistent with
negative infinity in the log.
Likewise, contours with no YSOs are plotted as if they had one YSO,
again with only upward uncertainties, assuming an uncertainty of 1
in the number of YSOs.
The blue line is the prediction of equation \ref{voltheoryeq},
translated to surface densities.
The black line is the result of a least-squares fit to the data.
}
\label{sdtI}
\end{figure}

\begin{figure}
\center
\includegraphics[scale=0.5, angle=-90]{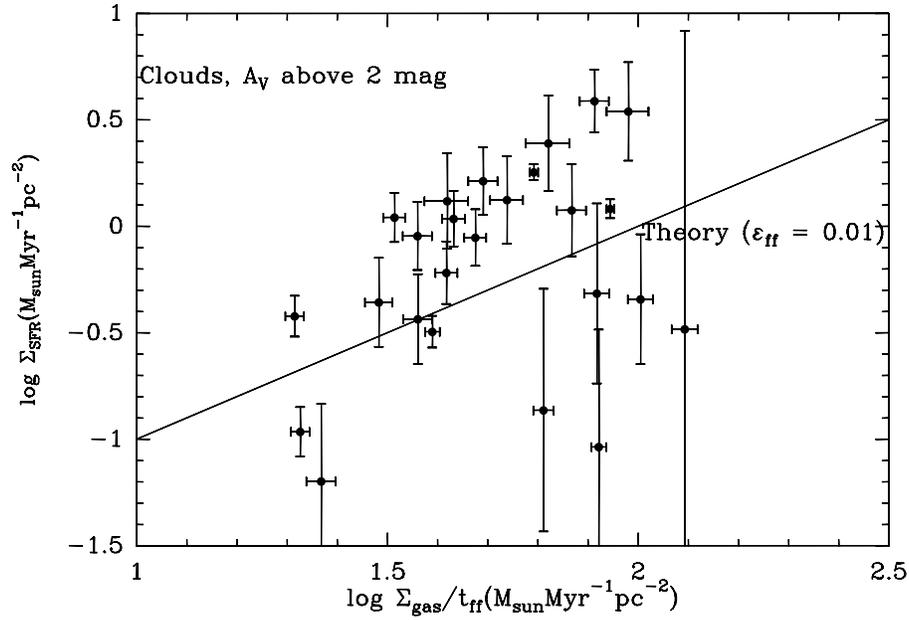}
\caption{
Plot of log(\sigmasfr) versus log(\sigmagas/\tff) for all YSOs in the
nearby clouds, on a cloud by cloud basis.   
Only clouds with at least one YSO are plotted.
The black line is the prediction of equation \ref{voltheoryeq}.
Statistical tests indicate no convincing correlation.
}
\label{cloudtffav2}
\end{figure}

\begin{figure}
\center
\includegraphics[scale=0.5, angle=-90]{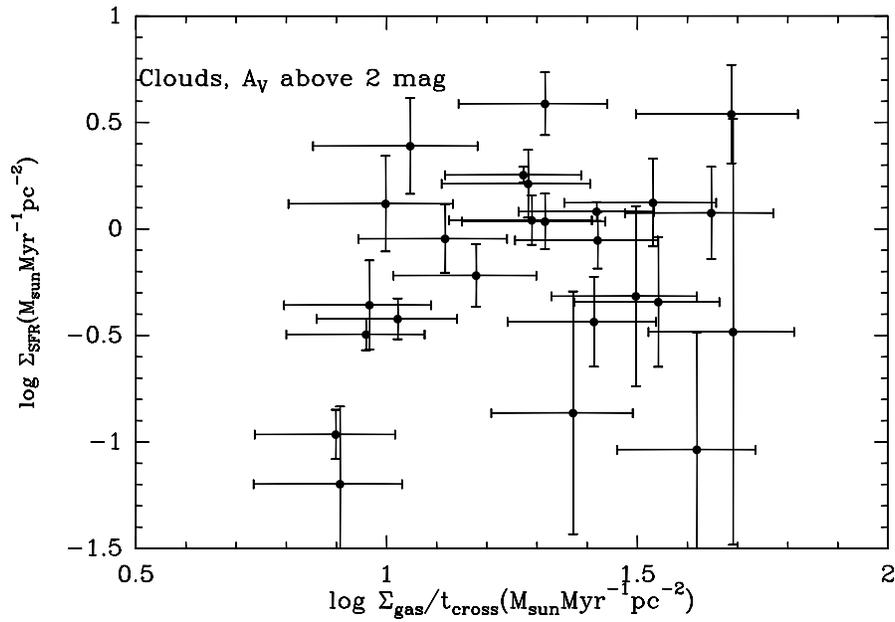}
\caption{
Plot of log(\sigmasfr) versus log(\sigmagas/\tcross) for all YSOs in the
nearby clouds.   
Statistical tests indicate no convincing correlation.
}
\label{tcross}
\end{figure}

\begin{figure}
\center
\includegraphics[scale=0.5, angle=0]{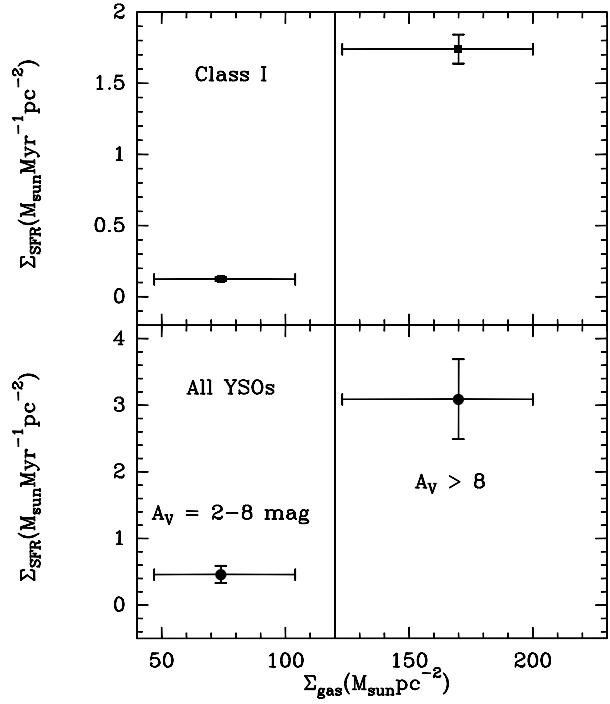}
\caption{
Plot of mean values of \sigmasfr\ versus \sigmagas\ for all YSOs in the
nearby clouds (lower).  In the upper panel, only the Class I
YSOs are included. The vertical line is at $\sigmagas = 120$
\msunpc, with the conversion from \av\ to \sigmagas\ adopted here.
}
\label{thresh}
\end{figure}

\begin{figure}
\center
\includegraphics[scale=0.5, angle=-90]{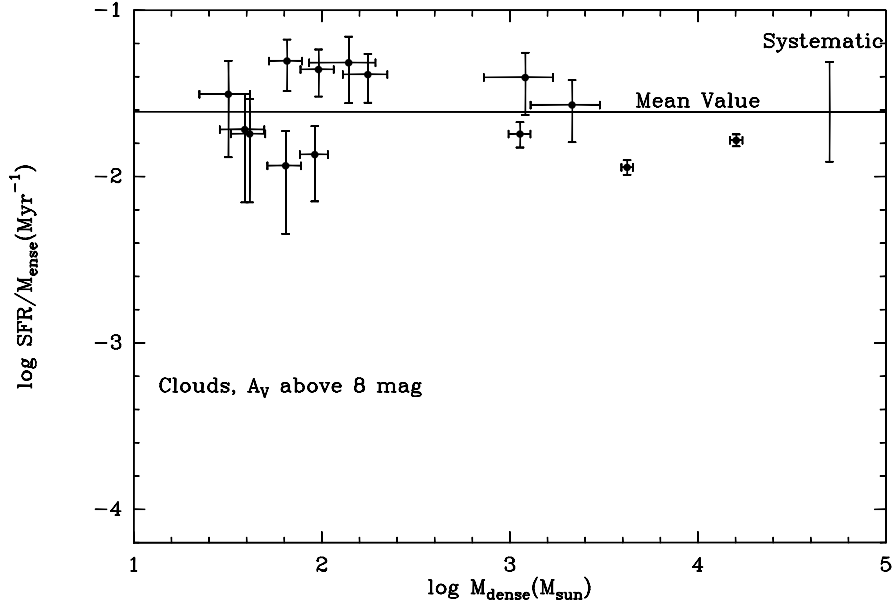}
\caption{
Plot of the logarithm of the SFR per mass of dense gas versus the 
logarithm of the mass of dense gas. 
The points are
from the nearby clouds counting YSOs and mass at $\av \ge 8$ mag.
The line is the mean value, and the error bars at 4.5 on the abscissa
indicate plausible uncertainties of the individual points. 
The plot scale is the same as for Fig. \ref{leecloud}.
}
\label{leeidea}
\end{figure}

\begin{figure}
\center
\includegraphics[scale=0.5, angle=-90]{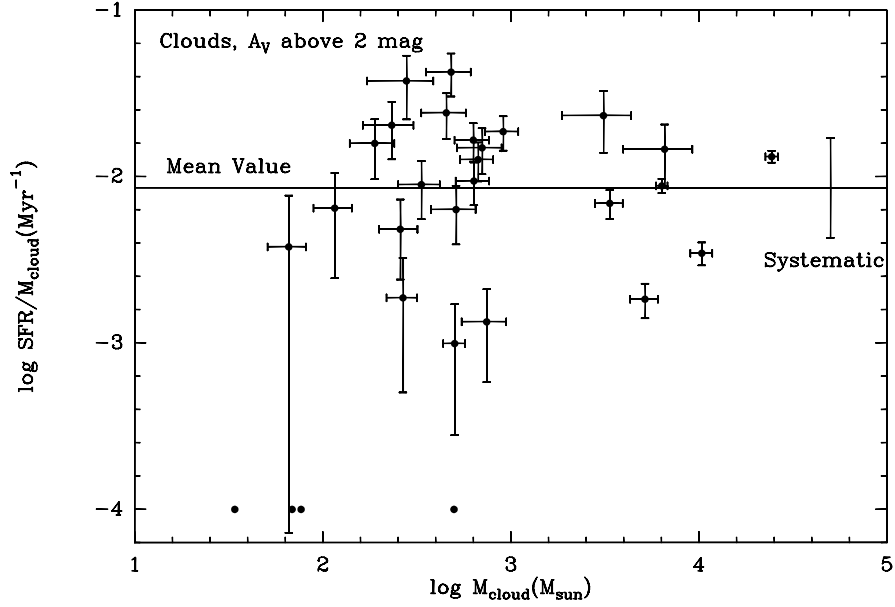}
\caption{
Plot of the logarithm of the SFR per total mass of the cloud versus the 
logarithm of the total cloud mass.
The points are from the nearby clouds counting YSOs. The points plotted
at -4.0 on the y-axis have no YSOs, and their location on the y axis
is arbitrary.
The line is the mean value, and the error bars at 4.5 on the abscissa
indicate plausible systematic uncertainties.
}
\label{leecloud}
\end{figure}

\begin{figure}
\center
\includegraphics[scale=0.5, angle=-90]{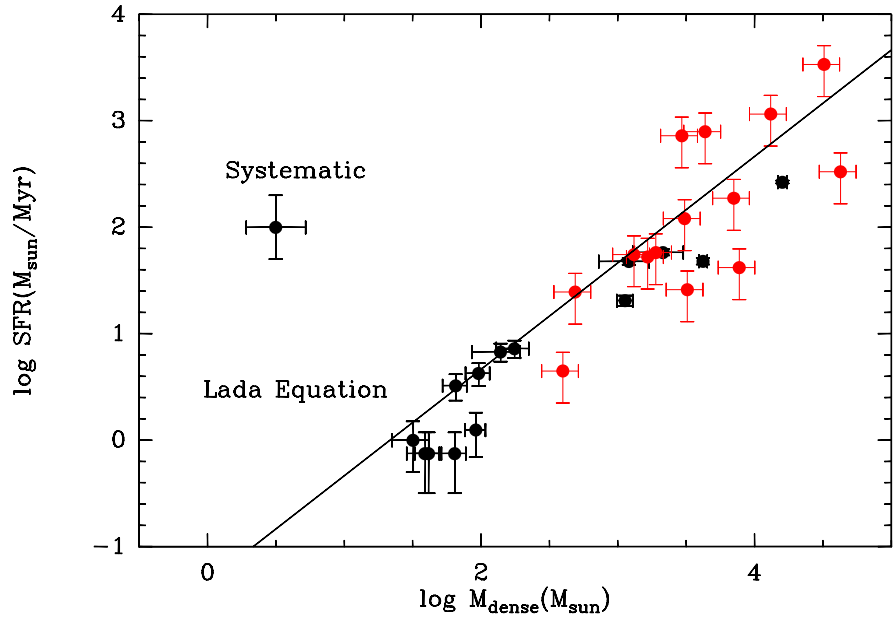}
\caption{
Plot of the logarithm of the SFR in the dense gas versus the 
logarithm of the mass of dense gas. The black points are
from the nearby clouds counting YSOs and mass at $\av \ge 8$ mag.
The red points are from massive dense clumps with SFR from radio
continuum and mass of dense gas as the virial mass measured by 
HCN \jj10\ emission.
The line is the prediction from \citet{2012ApJ...745..190L}.
}
\label{ladatesthigh}
\end{figure}

%% file: preprint.bbl
\begin{thebibliography}{84}
\expandafter\ifx\csname natexlab\endcsname\relax\def\natexlab#1{#1}\fi

\bibitem[{{Andr{\'e}} {et~al.}(2010){Andr{\'e}}, {Men'shchikov}, {Bontemps},
  {K{\"o}nyves}, {Motte}, {Schneider}, {Didelon}, {Minier}, {Saraceno},
  {Ward-Thompson}, {di Francesco}, {White}, {Molinari}, {Testi}, {Abergel},
  {Griffin}, {Henning}, {Royer}, {Mer{\'{\i}}n}, {Vavrek}, {Attard},
  {Arzoumanian}, {Wilson}, {Ade}, {Aussel}, {Baluteau}, {Benedettini},
  {Bernard}, {Blommaert}, {Cambr{\'e}sy}, {Cox}, {di Giorgio}, {Hargrave},
  {Hennemann}, {Huang}, {Kirk}, {Krause}, {Launhardt}, {Leeks}, {Le Pennec},
  {Li}, {Martin}, {Maury}, {Olofsson}, {Omont}, {Peretto}, {Pezzuto}, {Prusti},
  {Roussel}, {Russeil}, {Sauvage}, {Sibthorpe}, {Sicilia-Aguilar}, {Spinoglio},
  {Waelkens}, {Woodcraft}, \& {Zavagno}}]{2010A&A...518L.102A}
{Andr{\'e}}, P., {Men'shchikov}, A., {Bontemps}, S., {et~al.} 2010, \aap, 518,
  L102

\bibitem[{{Andr{\'e}} {et~al.}(2011){Andr{\'e}}, {Men'shchikov}, {Koenyves},
  {Arzoumanian}, {Peretto}, \& {Palmeirim}}]{2011sca..conf..321A}
{Andr{\'e}}, P., {Men'shchikov}, A., {Koenyves}, V., {et~al.} 2011, in Stellar
  Clusters {\amp} Associations: A RIA Workshop on Gaia, 321--328

\bibitem[{{Arzoumanian} {et~al.}(2011){Arzoumanian}, {Andr{\'e}}, {Didelon},
  {K{\"o}nyves}, {Schneider}, {Men'shchikov}, {Sousbie}, {Zavagno}, {Bontemps},
  {di Francesco}, {Griffin}, {Hennemann}, {Hill}, {Kirk}, {Martin}, {Minier},
  {Molinari}, {Motte}, {Peretto}, {Pezzuto}, {Spinoglio}, {Ward-Thompson},
  {White}, \& {Wilson}}]{2011A&A...529L...6A}
{Arzoumanian}, D., {Andr{\'e}}, P., {Didelon}, P., {et~al.} 2011, \aap, 529, L6

\bibitem[{{Arzoumanian} {et~al.}(2013){Arzoumanian}, {Andr{\'e}}, {Peretto}, \&
  {K{\"o}nyves}}]{2013A&A...553A.119A}
{Arzoumanian}, D., {Andr{\'e}}, P., {Peretto}, N., \& {K{\"o}nyves}, V. 2013,
  \aap, 553, A119

\bibitem[{{Ascenso} {et~al.}(2013){Ascenso}, {Lada}, {Alves},
  {Rom{\'a}n-Z{\'u}{\~n}iga}, \& {Lombardi}}]{2013A&A...549A.135A}
{Ascenso}, J., {Lada}, C.~J., {Alves}, J., {Rom{\'a}n-Z{\'u}{\~n}iga}, C.~G.,
  \& {Lombardi}, M. 2013, \aap, 549, A135

\bibitem[{{Bigiel} {et~al.}(2008){Bigiel}, {Leroy}, {Walter}, {Brinks}, {de
  Blok}, {Madore}, \& {Thornley}}]{2008AJ....136.2846B}
{Bigiel}, F., {Leroy}, A., {Walter}, F., {et~al.} 2008, \aj, 136, 2846

\bibitem[{{Bolatto} {et~al.}(2013){Bolatto}, {Wolfire}, \&
  {Leroy}}]{2013ARA&A..51..207B}
{Bolatto}, A.~D., {Wolfire}, M., \& {Leroy}, A.~K. 2013, \araa, 51, 207

\bibitem[{{Burkert} \& {Hartmann}(2013)}]{2013ApJ...773...48B}
{Burkert}, A. \& {Hartmann}, L. 2013, \apj, 773, 48

\bibitem[{{Chabrier}(2003)}]{2003PASP..115..763C}
{Chabrier}, G. 2003, \pasp, 115, 763

\bibitem[{{Chapman} {et~al.}(2009){Chapman}, {Mundy}, {Lai}, \&
  {Evans}}]{2009ApJ...690..496C}
{Chapman}, N.~L., {Mundy}, L.~G., {Lai}, S.-P., \& {Evans}, II, N.~J. 2009,
  \apj, 690, 496

\bibitem[{{Clark} \& {Glover}(2013)}]{2013arXiv1306.5714C}
{Clark}, P.~C. \& {Glover}, S.~C.~O. 2013, ArXiv e-prints

\bibitem[{{Daddi} {et~al.}(2010){Daddi}, {Elbaz}, {Walter}, {Bournaud},
  {Salmi}, {Carilli}, {Dannerbauer}, {Dickinson}, {Monaco}, \&
  {Riechers}}]{2010ApJ...714L.118D}
{Daddi}, E., {Elbaz}, D., {Walter}, F., {et~al.} 2010, \apjl, 714, L118

\bibitem[{{Dobashi} {et~al.}(1992){Dobashi}, {Yonekura}, {Mizuno}, \&
  {Fukui}}]{1992AJ....104.1525D}
{Dobashi}, K., {Yonekura}, Y., {Mizuno}, A., \& {Fukui}, Y. 1992, \aj, 104,
  1525

\bibitem[{{Dunham} {et~al.}(2013){Dunham}, {Arce}, {Allen}, {Evans},
  {Broekhoven-Fiene}, {Chapman}, {Cieza}, {Gutermuth}, {Harvey}, {Hatchell},
  {Huard}, {Kirk}, {Matthews}, {Mer{\'{\i}}n}, {Miller}, {Peterson}, \&
  {Spezzi}}]{2013AJ....145...94D}
{Dunham}, M.~M., {Arce}, H.~G., {Allen}, L.~E., {et~al.} 2013, \aj, 145, 94

\bibitem[{{Dzib} {et~al.}(2011){Dzib}, {Loinard}, {Mioduszewski}, {Boden},
  {Rodr{\'{\i}}guez}, \& {Torres}}]{2011RMxAC..40..231D}
{Dzib}, S., {Loinard}, L., {Mioduszewski}, A.~J., {et~al.} 2011, in Revista
  Mexicana de Astronomia y Astrofisica Conference Series, Vol.~40, Revista
  Mexicana de Astronomia y Astrofisica Conference Series, 231--232

\bibitem[{{Elmegreen}(2000)}]{2000ApJ...530..277E}
{Elmegreen}, B.~G. 2000, \apj, 530, 277

\bibitem[{{Enoch} {et~al.}(2009){Enoch}, {Evans}, {Sargent}, \&
  {Glenn}}]{2009ApJ...692..973E}
{Enoch}, M.~L., {Evans}, II, N.~J., {Sargent}, A.~I., \& {Glenn}, J. 2009,
  \apj, 692, 973

\bibitem[{{Enoch} {et~al.}(2008){Enoch}, {Evans}, {Sargent}, {Glenn},
  {Rosolowsky}, \& {Myers}}]{2008ApJ...684.1240E}
{Enoch}, M.~L., {Evans}, II, N.~J., {Sargent}, A.~I., {et~al.} 2008, \apj, 684,
  1240

\bibitem[{{Enoch} {et~al.}(2007){Enoch}, {Glenn}, {Evans}, {Sargent}, {Young},
  \& {Huard}}]{2007ApJ...666..982E}
{Enoch}, M.~L., {Glenn}, J., {Evans}, II, N.~J., {et~al.} 2007, \apj, 666, 982

\bibitem[{{Evans}(2007)}]{evans07}
{Evans}, N.~J., {\rm et. al.}. 2007, Final Delivery of Data from the c2d Legacy
  Project: IRAC and MIPS (Pasadena: SSC),
  \url{http://ssc.spitzer.caltech.edu/legacy/all.html}

\bibitem[{{Evans} {et~al.}(2003){Evans}, {Allen}, {Blake}, {Boogert}, {Bourke},
  {Harvey}, {Kessler}, {Koerner}, {Lee}, {Mundy}, {Myers}, {Padgett},
  {Pontoppidan}, {Sargent}, {Stapelfeldt}, {van Dishoeck}, {Young}, \&
  {Young}}]{2003PASP..115..965E}
{Evans}, II, N.~J., {Allen}, L.~E., {Blake}, G.~A., {et~al.} 2003, \pasp, 115,
  965

\bibitem[{{Evans} {et~al.}(2009){Evans}, {Dunham}, {J{\o}rgensen}, {Enoch},
  {Mer{\'{\i}}n}, {van Dishoeck}, {Alcal{\'a}}, {Myers}, {Stapelfeldt},
  {Huard}, {Allen}, {Harvey}, {van Kempen}, {Blake}, {Koerner}, {Mundy},
  {Padgett}, \& {Sargent}}]{2009ApJS..181..321E}
{Evans}, II, N.~J., {Dunham}, M.~M., {J{\o}rgensen}, J.~K., {et~al.} 2009,
  \apjs, 181, 321

\bibitem[{{Federrath} \& {Klessen}(2012)}]{2012ApJ...761..156F}
{Federrath}, C. \& {Klessen}, R.~S. 2012, \apj, 761, 156

\bibitem[{{Gao} \& {Solomon}(2004)}]{2004ApJ...606..271G}
{Gao}, Y. \& {Solomon}, P.~M. 2004, \apj, 606, 271

\bibitem[{{Genzel} {et~al.}(2010){Genzel}, {Tacconi}, {Gracia-Carpio},
  {Sternberg}, {Cooper}, {Shapiro}, {Bolatto}, {Bouch{\'e}}, {Bournaud},
  {Burkert}, {Combes}, {Comerford}, {Cox}, {Davis}, {Schreiber},
  {Garcia-Burillo}, {Lutz}, {Naab}, {Neri}, {Omont}, {Shapley}, \&
  {Weiner}}]{2010MNRAS.407.2091G}
{Genzel}, R., {Tacconi}, L.~J., {Gracia-Carpio}, J., {et~al.} 2010, \mnras,
  407, 2091

\bibitem[{{Goldsmith} {et~al.}(2008){Goldsmith}, {Heyer}, {Narayanan}, {Snell},
  {Li}, \& {Brunt}}]{2008ApJ...680..428G}
{Goldsmith}, P.~F., {Heyer}, M., {Narayanan}, G., {et~al.} 2008, \apj, 680, 428

\bibitem[{{Gutermuth} {et~al.}(2008){Gutermuth}, {Bourke}, {Allen}, {Myers},
  {Megeath}, {Matthews}, {J{\o}rgensen}, {Di Francesco}, {Ward-Thompson},
  {Huard}, {Brooke}, {Dunham}, {Cieza}, {Harvey}, \&
  {Chapman}}]{2008ApJ...673L.151G}
{Gutermuth}, R.~A., {Bourke}, T.~L., {Allen}, L.~E., {et~al.} 2008, \apjl, 673,
  L151

\bibitem[{{Gutermuth} {et~al.}(2011){Gutermuth}, {Pipher}, {Megeath}, {Myers},
  {Allen}, \& {Allen}}]{2011ApJ...739...84G}
{Gutermuth}, R.~A., {Pipher}, J.~L., {Megeath}, S.~T., {et~al.} 2011, \apj,
  739, 84

\bibitem[{{Hara} {et~al.}(1999){Hara}, {Tachihara}, {Mizuno}, {Onishi},
  {Kawamura}, {Obayashi}, \& {Fukui}}]{1999PASJ...51..895H}
{Hara}, A., {Tachihara}, K., {Mizuno}, A., {et~al.} 1999, \pasj, 51, 895

\bibitem[{{Harvey} {et~al.}(2007){Harvey}, {Mer{\'{\i}}n}, {Huard}, {Rebull},
  {Chapman}, {Evans}, \& {Myers}}]{2007ApJ...663.1149H}
{Harvey}, P., {Mer{\'{\i}}n}, B., {Huard}, T.~L., {et~al.} 2007, \apj, 663,
  1149

\bibitem[{{Harvey} {et~al.}(2013){Harvey}, {Fallscheer}, {Ginsburg}, {Terebey},
  {Andr{\'e}}, {Bourke}, {Di Francesco}, {K{\"o}nyves}, {Matthews}, \&
  {Peterson}}]{2013ApJ...764..133H}
{Harvey}, P.~M., {Fallscheer}, C., {Ginsburg}, A., {et~al.} 2013, \apj, 764,
  133

\bibitem[{{Hatchell} {et~al.}(2012){Hatchell}, {Terebey}, {Huard}, {Mamajek},
  {Allen}, {Bourke}, {Dunham}, {Gutermuth}, {Harvey}, {J{\o}rgensen},
  {Mer{\'{\i}}n}, {Noriega-Crespo}, \& {Peterson}}]{2012ApJ...754..104H}
{Hatchell}, J., {Terebey}, S., {Huard}, T., {et~al.} 2012, \apj, 754, 104

\bibitem[{{Heiderman} {et~al.}(2010){Heiderman}, {Evans}, {Allen}, {Huard}, \&
  {Heyer}}]{2010ApJ...723.1019H}
{Heiderman}, A., {Evans}, II, N.~J., {Allen}, L.~E., {Huard}, T., \& {Heyer},
  M. 2010, \apj, 723, 1019

\bibitem[{{Hennebelle} \& {Chabrier}(2011)}]{2011ApJ...743L..29H}
{Hennebelle}, P. \& {Chabrier}, G. 2011, \apjl, 743, L29

\bibitem[{{Herbertz} {et~al.}(1991){Herbertz}, {Ungerechts}, \&
  {Winnewisser}}]{1991A&A...249..483H}
{Herbertz}, R., {Ungerechts}, H., \& {Winnewisser}, G. 1991, \aap, 249, 483

\bibitem[{{Hsieh} \& {Lai}(2013)}]{2013ApJS..205....5H}
{Hsieh}, T.-H. \& {Lai}, S.-P. 2013, \apjs, 205, 5

\bibitem[{{Johnstone} {et~al.}(2004){Johnstone}, {Di Francesco}, \&
  {Kirk}}]{2004ApJ...611L..45J}
{Johnstone}, D., {Di Francesco}, J., \& {Kirk}, H. 2004, \apjl, 611, L45

\bibitem[{{Kennicutt} \& {Evans}(2012)}]{2012ARA&A..50..531K}
{Kennicutt}, R.~C. \& {Evans}, N.~J. 2012, \araa, 50, 531

\bibitem[{{Kennicutt}(1998)}]{1998ApJ...498..541K}
{Kennicutt}, Jr., R.~C. 1998, \apj, 498, 541

\bibitem[{{Kirk} {et~al.}(2009){Kirk}, {Ward-Thompson}, {Di Francesco},
  {Bourke}, {Evans}, {Mer{\'{\i}}n}, {Allen}, {Cieza}, {Dunham}, {Harvey},
  {Huard}, {J{\o}rgensen}, {Miller}, {Noriega-Crespo}, {Peterson}, {Ray}, \&
  {Rebull}}]{2009ApJS..185..198K}
{Kirk}, J.~M., {Ward-Thompson}, D., {Di Francesco}, J., {et~al.} 2009, \apjs,
  185, 198

\bibitem[{{Knapp} {et~al.}(1976){Knapp}, {Kuiper}, {Knapp}, \&
  {Brown}}]{1976ApJ...206..443K}
{Knapp}, G.~R., {Kuiper}, T.~B.~H., {Knapp}, S.~L., \& {Brown}, R.~L. 1976,
  \apj, 206, 443

\bibitem[{{Kroupa}(2002)}]{2002Sci...295...82K}
{Kroupa}, P. 2002, Science, 295, 82

\bibitem[{{Kruijssen} {et~al.}(2013){Kruijssen}, {Longmore}, {Elmegreen},
  {Murray}, {Bally}, {Testi}, \& {Kennicutt}}]{2013arXiv1303.6286K}
{Kruijssen}, J.~M.~D., {Longmore}, S.~N., {Elmegreen}, B.~G., {et~al.} 2013,
  ArXiv e-prints

\bibitem[{{Krumholz} {et~al.}(2012){Krumholz}, {Dekel}, \&
  {McKee}}]{2012ApJ...745...69K}
{Krumholz}, M.~R., {Dekel}, A., \& {McKee}, C.~F. 2012, \apj, 745, 69

\bibitem[{{Krumholz} \& {McKee}(2005)}]{2005ApJ...630..250K}
{Krumholz}, M.~R. \& {McKee}, C.~F. 2005, \apj, 630, 250

\bibitem[{{Krumholz} \& {Thompson}(2007)}]{2007ApJ...669..289K}
{Krumholz}, M.~R. \& {Thompson}, T.~A. 2007, \apj, 669, 289

\bibitem[{{Lada} {et~al.}(2012){Lada}, {Forbrich}, {Lombardi}, \&
  {Alves}}]{2012ApJ...745..190L}
{Lada}, C.~J., {Forbrich}, J., {Lombardi}, M., \& {Alves}, J.~F. 2012, \apj,
  745, 190

\bibitem[{{Lada} {et~al.}(2009){Lada}, {Lombardi}, \&
  {Alves}}]{2009ApJ...703...52L}
{Lada}, C.~J., {Lombardi}, M., \& {Alves}, J.~F. 2009, \apj, 703, 52

\bibitem[{{Lada} {et~al.}(2010){Lada}, {Lombardi}, \&
  {Alves}}]{2010ApJ...724..687L}
{Lada}, C.~J., {Lombardi}, M., \& {Alves}, J.~F. 2010, \apj, 724, 687

\bibitem[{{Lada} {et~al.}(2013){Lada}, {Lombardi}, {Roman-Zuniga}, {Forbrich},
  \& {Alves}}]{2013ApJ...778..133L}
{Lada}, C.~J., {Lombardi}, M., {Roman-Zuniga}, C., {Forbrich}, J., \& {Alves},
  J.~F. 2013, \apj, 778, 133

\bibitem[{{Lada}(1992)}]{1992ApJ...393L..25L}
{Lada}, E.~A. 1992, \apjl, 393, L25

\bibitem[{{Leroy} {et~al.}(2008){Leroy}, {Walter}, {Brinks}, {Bigiel}, {de
  Blok}, {Madore}, \& {Thornley}}]{2008AJ....136.2782L}
{Leroy}, A.~K., {Walter}, F., {Brinks}, E., {et~al.} 2008, \aj, 136, 2782

\bibitem[{{Leroy} {et~al.}(2013){Leroy}, {Walter}, {Sandstrom}, {Schruba},
  {Munoz-Mateos}, {Bigiel}, {Bolatto}, {Brinks}, {de Blok}, {Meidt}, {Rix},
  {Rosolowsky}, {Schinnerer}, {Schuster}, \& {Usero}}]{2013AJ....146...19L}
{Leroy}, A.~K., {Walter}, F., {Sandstrom}, K., {et~al.} 2013, \aj, 146, 19

\bibitem[{{Li} {et~al.}(1997){Li}, {Evans}, \& {Lada}}]{1997ApJ...488..277L}
{Li}, W., {Evans}, II, N.~J., \& {Lada}, E.~A. 1997, \apj, 488, 277

\bibitem[{{Longmore} {et~al.}(2013){Longmore}, {Bally}, {Testi}, {Purcell},
  {Walsh}, {Bressert}, {Pestalozzi}, {Molinari}, {Ott}, {Cortese}, {Battersby},
  {Murray}, {Lee}, {Kruijssen}, {Schisano}, \& {Elia}}]{2013MNRAS.429..987L}
{Longmore}, S.~N., {Bally}, J., {Testi}, L., {et~al.} 2013, \mnras, 429, 987

\bibitem[{{Loren}(1979)}]{1979ApJ...227..832L}
{Loren}, R.~B. 1979, \apj, 227, 832

\bibitem[{{Maury} {et~al.}(2011){Maury}, {Andr{\'e}}, {Men'shchikov},
  {K{\"o}nyves}, \& {Bontemps}}]{2011A&A...535A..77M}
{Maury}, A.~J., {Andr{\'e}}, P., {Men'shchikov}, A., {K{\"o}nyves}, V., \&
  {Bontemps}, S. 2011, \aap, 535, A77

\bibitem[{{McKee}(1989)}]{1989ApJ...345..782M}
{McKee}, C.~F. 1989, \apj, 345, 782

\bibitem[{{Mizuno} {et~al.}(1998){Mizuno}, {Hayakawa}, {Yamaguchi}, {Kato},
  {Hara}, {Mizuno}, {Yonekura}, {Onishi}, {Kawamura}, {Tachihara}, {Obayashi},
  {Xiao}, {Ogawa}, \& {Fukui}}]{1998ApJ...507L..83M}
{Mizuno}, A., {Hayakawa}, T., {Yamaguchi}, N., {et~al.} 1998, \apjl, 507, L83

\bibitem[{{Mizuno} {et~al.}(2001){Mizuno}, {Yamaguchi}, {Tachihara}, {Toyoda},
  {Aoyama}, {Yamamoto}, {Onishi}, \& {Fukui}}]{2001PASJ...53.1071M}
{Mizuno}, A., {Yamaguchi}, R., {Tachihara}, K., {et~al.} 2001, \pasj, 53, 1071

\bibitem[{{Mouschovias} \& {Spitzer}(1976)}]{1976ApJ...210..326M}
{Mouschovias}, T.~C. \& {Spitzer}, Jr., L. 1976, \apj, 210, 326

\bibitem[{{Nakamura} {et~al.}(2011){Nakamura}, {Sugitani}, {Shimajiri},
  {Tsukagoshi}, {Higuchi}, {Nishiyama}, {Kawabe}, {Takami}, {Karr},
  {Gutermuth}, \& {Wilson}}]{2011ApJ...737...56N}
{Nakamura}, F., {Sugitani}, K., {Shimajiri}, Y., {et~al.} 2011, \apj, 737, 56

\bibitem[{{Narayanan} {et~al.}(2008){Narayanan}, {Cox}, {Shirley}, {Dav{\'e}},
  {Hernquist}, \& {Walker}}]{2008ApJ...684..996N}
{Narayanan}, D., {Cox}, T.~J., {Shirley}, Y., {et~al.} 2008, \apj, 684, 996

\bibitem[{{Ninkovic} \& {Trajkovska}(2006)}]{2006SerAJ.172...17N}
{Ninkovic}, S. \& {Trajkovska}, V. 2006, Serbian Astronomical Journal, 172, 17

\bibitem[{{Nozawa} {et~al.}(1991){Nozawa}, {Mizuno}, {Teshima}, {Ogawa}, \&
  {Fukui}}]{1991ApJS...77..647N}
{Nozawa}, S., {Mizuno}, A., {Teshima}, Y., {Ogawa}, H., \& {Fukui}, Y. 1991,
  \apjs, 77, 647

\bibitem[{{Oliveira} {et~al.}(2009){Oliveira}, {Mer{\'{\i}}n}, {Pontoppidan},
  {van Dishoeck}, {Overzier}, {Hern{\'a}ndez}, {Sicilia-Aguilar}, {Eiroa}, \&
  {Montesinos}}]{2009ApJ...691..672O}
{Oliveira}, I., {Mer{\'{\i}}n}, B., {Pontoppidan}, K.~M., {et~al.} 2009, \apj,
  691, 672

\bibitem[{{Onishi} {et~al.}(1998){Onishi}, {Mizuno}, {Kawamura}, {Ogawa}, \&
  {Fukui}}]{1998ApJ...502..296O}
{Onishi}, T., {Mizuno}, A., {Kawamura}, A., {Ogawa}, H., \& {Fukui}, Y. 1998,
  \apj, 502, 296

\bibitem[{{Padoan} {et~al.}(2013){Padoan}, {Federrath}, {Chabrier}, {Evans},
  {Johnstone}, {J{\o}rgensen}, {McKee}, \& {Nordlund}}]{2013arXiv1312.5365P}
{Padoan}, P., {Federrath}, C., {Chabrier}, G., {et~al.} 2013, ArXiv e-prints

\bibitem[{{Press} {et~al.}(1992){Press}, {Teukolsky}, {Vetterling}, \&
  {Flannery}}]{1992nrfa.book.....P}
{Press}, W.~H., {Teukolsky}, S.~A., {Vetterling}, W.~T., \& {Flannery}, B.~P.
  1992, {Numerical recipes in FORTRAN. The art of scientific computing}

\bibitem[{{Ridge} {et~al.}(2006){Ridge}, {Di Francesco}, {Kirk}, {Li},
  {Goodman}, {Alves}, {Arce}, {Borkin}, {Caselli}, {Foster}, {Heyer},
  {Johnstone}, {Kosslyn}, {Lombardi}, {Pineda}, {Schnee}, \&
  {Tafalla}}]{2006AJ....131.2921R}
{Ridge}, N.~A., {Di Francesco}, J., {Kirk}, H., {et~al.} 2006, \aj, 131, 2921

\bibitem[{{Schmidt}(1959)}]{1959ApJ...129..243S}
{Schmidt}, M. 1959, \apj, 129, 243

\bibitem[{{Schmidt}(1963)}]{1963ApJ...137..758S}
{Schmidt}, M. 1963, \apj, 137, 758

\bibitem[{{Schruba} {et~al.}(2011){Schruba}, {Leroy}, {Walter}, {Bigiel},
  {Brinks}, {de Blok}, {Dumas}, {Kramer}, {Rosolowsky}, {Sandstrom},
  {Schuster}, {Usero}, {Weiss}, \& {Wiesemeyer}}]{2011AJ....142...37S}
{Schruba}, A., {Leroy}, A.~K., {Walter}, F., {et~al.} 2011, \aj, 142, 37

\bibitem[{{Snell} {et~al.}(2002){Snell}, {Carpenter}, \&
  {Heyer}}]{2002ApJ...578..229S}
{Snell}, R.~L., {Carpenter}, J.~M., \& {Heyer}, M.~H. 2002, \apj, 578, 229

\bibitem[{{Solomon} \& {Sage}(1988)}]{1988ApJ...334..613S}
{Solomon}, P.~M. \& {Sage}, L.~J. 1988, \apj, 334, 613

\bibitem[{{Spezzi} {et~al.}(2008){Spezzi}, {Alcal{\'a}}, {Covino}, {Frasca},
  {Gandolfi}, {Oliveira}, {Chapman}, {Evans}, {Huard}, {J{\o}rgensen},
  {Mer{\'{\i}}n}, \& {Stapelfeldt}}]{2008ApJ...680.1295S}
{Spezzi}, L., {Alcal{\'a}}, J.~M., {Covino}, E., {et~al.} 2008, \apj, 680, 1295

\bibitem[{{Tothill} {et~al.}(2009){Tothill}, {L{\"o}hr}, {Parshley}, {Stark},
  {Lane}, {Harnett}, {Wright}, {Walker}, {Bourke}, \&
  {Myers}}]{2009ApJS..185...98T}
{Tothill}, N.~F.~H., {L{\"o}hr}, A., {Parshley}, S.~C., {et~al.} 2009, \apjs,
  185, 98

\bibitem[{{van Kempen} {et~al.}(2009){van Kempen}, {van Dishoeck}, {Salter},
  {Hogerheijde}, {J{\o}rgensen}, \& {Boogert}}]{2009A&A...498..167V}
{van Kempen}, T.~A., {van Dishoeck}, E.~F., {Salter}, D.~M., {et~al.} 2009,
  \aap, 498, 167

\bibitem[{{Vilas-Boas} {et~al.}(1994){Vilas-Boas}, {Myers}, \&
  {Fuller}}]{1994ApJ...433...96V}
{Vilas-Boas}, J.~W.~S., {Myers}, P.~C., \& {Fuller}, G.~A. 1994, \apj, 433, 96

\bibitem[{{Vutisalchavakul} \& {Evans}(2013)}]{2013ApJ...765..129V}
{Vutisalchavakul}, N. \& {Evans}, II, N.~J. 2013, \apj, 765, 129

\bibitem[{{Weingartner} \& {Draine}(2001)}]{2001ApJ...548..296W}
{Weingartner}, J.~C. \& {Draine}, B.~T. 2001, \apj, 548, 296

\bibitem[{{Wu} {et~al.}(2005){Wu}, {Evans}, {Gao}, {Solomon}, {Shirley}, \&
  {Vanden Bout}}]{2005ApJ...635L.173W}
{Wu}, J., {Evans}, II, N.~J., {Gao}, Y., {et~al.} 2005, \apjl, 635, L173

\bibitem[{{Wu} {et~al.}(2010){Wu}, {Evans}, {Shirley}, \&
  {Knez}}]{2010ApJS..188..313W}
{Wu}, J., {Evans}, II, N.~J., {Shirley}, Y.~L., \& {Knez}, C. 2010, \apjs, 188,
  313

\bibitem[{{Zuckerman} \& {Evans}(1974)}]{1974ApJ...192L.149Z}
{Zuckerman}, B. \& {Evans}, II, N.~J. 1974, \apjl, 192, L149

\end{thebibliography}
